\documentclass[twocolumn]{aastex62}
\usepackage{amsmath}
\usepackage{amssymb}
\usepackage{mathtools}
\usepackage{multirow}

\providecommand{\deriv}{\ensuremath{\mathrm{d}}}
\providecommand{\given}{\ensuremath{\hspace{0.05em}\mid\hspace{0.05em}}}
\providecommand{\logten}{\ensuremath{\log_{\rm 10}}}

\providecommand{\deriv}{\ensuremath{\mathrm{d}}}
\providecommand{\given}{\ensuremath{\hspace{0.01em}\mid\hspace{0.01em}}}
\providecommand{\logten}{\ensuremath{\log_{\rm 10}}}
\providecommand{\gaia}{Gaia}
\providecommand{\gdr}[1]{GDR{#1}}
\providecommand{\gedr}[1]{GeDR{#1}}
\providecommand{\mock}{GeDR3mock}
\providecommand{\release}{EDR3}
\providecommand{\gmag}{\ensuremath{G}}
\providecommand{\MG}{\ensuremath{M_G}}
\providecommand{\AG}{\ensuremath{A_G}}
\providecommand{\QG}{\ensuremath{Q_G}}
\providecommand{\bprp}{\ensuremath{{\rm BP} \! - \! {\rm RP}}}
\providecommand{\cc}{\ensuremath{c}}
\providecommand{\parallax}{\ensuremath{\varpi}}
\providecommand{\parallaxzp}{\ensuremath{\varpi_{\rm zp}}}
\providecommand{\sigparallax}{\ensuremath{\sigma_{\varpi}}}
\providecommand{\fpu}{\ensuremath{\sigparallax/\parallax}}
\providecommand{\glon}{\ensuremath{l}}
\providecommand{\glat}{\ensuremath{b}}
\providecommand{\hp}{\ensuremath{p}} 
\providecommand{\dist}{\ensuremath{r}} 
 
\providecommand{\rlo}{\ensuremath{\dist_{\rm lo}}} 
\providecommand{\rhi}{\ensuremath{\dist_{\rm hi}}} 
\providecommand{\rmode}{\ensuremath{\dist_{\rm mode}}}
\providecommand{\rmed}{\ensuremath{\dist_{\rm med}}}
\providecommand{\rEDSDmode}{\ensuremath{\dist^{\rm EDSD}_{\rm mode}}}
\providecommand{\rinit}{\ensuremath{\dist_{\rm init}}}
\providecommand{\rlen}{\ensuremath{L}}

\providecommand{\degree}{\ensuremath{^\circ}}

\received{9 December 2020}
\revised{30 December 2020}
\accepted{31 December 2020}
\journalinfo{Accepted to the Astronomical Journal} 

\shorttitle{\gaia\ \release\ distances}
\shortauthors{Bailer-Jones et al.}

\begin{document}


\title{Estimating distances from parallaxes. V:\\
Geometric and photogeometric distances to 1.47 billion stars in Gaia Early Data Release 3
}


\author{C.A.L.\ Bailer-Jones}
\affil{Max Planck Institute for Astronomy, Heidelberg, Germany}

\author{J.\ Rybizki}
\affil{Max Planck Institute for Astronomy, Heidelberg, Germany}

\author[0000-0001-9256-5516]{M.\ Fouesneau}
\affil{Max Planck Institute for Astronomy, Heidelberg, Germany}

\author{M.\ Demleitner}
\affil{Astronomisches Rechen-Institut, Zentrum f\"ur Astronomie der Universit\"at Heidelberg, Germany}

\author{R.\ Andrae}
\affil{Max Planck Institute for Astronomy, Heidelberg, Germany}

\begin{abstract}
Stellar distances constitute a foundational pillar of astrophysics. The publication of 1.47 billion stellar parallaxes from Gaia is a major contribution to this. Yet despite Gaia's precision, the majority of these stars are so distant or faint that their fractional parallax uncertainties are large, thereby precluding a simple inversion of parallax to provide a distance. Here we take a probabilistic approach to estimating stellar distances that uses a prior constructed from a three-dimensional model of our Galaxy. This model includes interstellar extinction and Gaia's variable magnitude limit. We infer two types of distance. The first, geometric, uses the parallax together with a direction-dependent prior on distance. The second, photogeometric, additionally uses the colour and apparent magnitude of a star, by exploiting the fact that stars of a given colour have a restricted range of probable absolute magnitudes (plus extinction). Tests on simulated data and external validations show that the photogeometric estimates generally have higher accuracy and precision for stars with poor parallaxes. We provide a catalogue of 1.47 billion geometric and 1.35 billion 
photogeometric distances together with asymmetric uncertainty measures. Our estimates are quantiles of a posterior probability distribution, so they transform invariably and can therefore also be used directly in the distance modulus ($5\logten\dist\,-\,5$). The catalogue may be downloaded or queried using ADQL at various sites (see \url{http://www.mpia.de/~calj/gedr3_distances.html}) where it can also be cross-matched with the Gaia catalogue.
\end{abstract}




\section{Introduction}\label{sec:introduction}

There are various ways to determine astrophysical distances. Near the base of the distance ladder on which almost all other distance measures are built are geometric parallaxes of stars. In recognition of this, the European Space Agency (ESA) implemented the \gaia\ mission to obtain parallaxes for over one billion stars in our Galaxy down to $\gmag \simeq 20$\,mag,
with accuracies to tens of microarcseconds \citep{2016A&A...595A...1G}. The first two data releases \citep{2016A&A...595A...2G, 2018A&A...616A...1G} presented a significant leap forward in both the number and accuracy of stellar parallaxes. The recently published early third release \citep{gedr3_release} (hereafter \release) reduces the random and systematic errors in the parallaxes by another 30\%.

While parallaxes (\parallax) are the basis for a distance determination, they are not themselves distances (\dist). 
This is due to the nonlinear transformation between them 
($\parallax \sim 1/\dist$)
and the presence of significant noise for more distant stars.  Small absolute uncertainties in parallax can translate into large uncertainties in distance, and while parallaxes can be negative, distances cannot be. 
Thus for anything but the most precise parallaxes, the inverse parallax is a poor distance estimate.
An explicit probabilistic approach to inferring distances may instead be taken.
This has been discussed and applied to parallax data in various publications in recent years; a recent overview is given by \cite{2018A&A...616A...9L}.
The simplest approach uses just the parallax and parallax uncertainty together with a one-dimensional prior over distance. This yields a posterior probability distribution over distance to an individual star \citep{2015PASP..127..994B}. A suitable prior ensures that the posterior converges to something sensible as the precision of the parallax degrades. This is important when working with \gaia\ data,
because its truly revolutionary nature notwithstanding, in \release\ 
43\% of the sources 
have parallax uncertainties greater than 50\% (63\% greater than 20\%), and a further 24\% have negative parallaxes.
The shape and scale of the prior distribution should reflect the expected distribution of stars in the sample, including observational selection effects such as magnitude limits. The prior's characteristic length scale will typically need to vary with direction in the Galaxy \citep{2018AJ....156...58B}.
More sophisticated approaches use other types of data,  
such as the star's magnitude and colour
\citep{2016ApJ...832..137A, 2018RNAAS...2...51M, 2019A&A...628A..94A, 2019MNRAS.489.2079L},
velocity \citep{2017MNRAS.472.3979S, 2018ApJ...869...83Z}, 
or spectroscopic \citep{2018MNRAS.481.4093S, 2020A&A...638A..76Q}
or asteroseismic \citep{2019MNRAS.486.3569H} parameters. In order to exploit such additional data, these
methods must make deeper astrophysical assumptions than parallax-only approaches, and may also have more complex priors. The benefit is that the inferred distances will usually be more precise (lower random errors), and hopefully also more accurate (lower systematic errors) if the extra assumptions are correct.

In the present paper, the fifth in a series, we determine distances for sources in \release\ using data exclusively from \release. The resulting catalogue should be more accurate and more useful than our earlier work, on account of both the more accurate parallaxes in \release\ and improvements in our method. We determine two types of distance. The first, which we call ``geometric", uses only the parallaxes and their uncertainties.
We explored this approach in detail in the first two papers in this series \citep{2015PASP..127..994B, 2016ApJ...832..137A} (hereafter papers I and II), and applied it to estimate distances for 2 million stars in the first \gaia\ data release
\citep{2016ApJ...833..119A} (paper~III)
and 1.33 billion stars in the second \gaia\ data release
\citep{2018AJ....156...58B} (paper~IV).
Both papers used a (different) direction-dependent distance prior that reflected the Galaxy's stellar populations and \gaia's selection thereof. 

Our second type of distance estimate uses, in addition to the parallax, the colour and magnitude of the star. We call such distances ``photogeometric". As well as the distance prior, this uses a model of the direction-dependent distribution of (extincted) stellar absolute magnitudes. 

We construct our priors from the \gedr{3} mock catalogue of \cite{2020PASP..132g4501R}.
This lists, among other things, the (noise-free) positions, distances, magnitudes, colours, and extinctions of 1.5 billion individual stars in the Galaxy 
as a mock-up of what was expected to appear in \release. \mock\ is based on the Besan\c{c}on Galactic model and PARSEC stellar evolutionary tracks.
We exclude stars from \mock\ that simulate the Magellanic Clouds ({\tt popid=10}) and stellar open clusters ({\tt popid=11}).
We divide the sky into the 12288 equal-area (3.36\,sq.~deg.)
regions defined by the HEALpixel scheme\footnote{\url{https://healpix.sourceforge.io}} at level 5, and fit our prior models separately to each. In doing this we only retain from \mock\ those stars that are brighter than the 90th percentile of the \release\ magnitude distribution in that HEALpixel
\citep{2018ascl.soft11018R, gedr3_nearbystars}.
This is done to mimic the variable magnitude limit of \gaia\ over the sky, and varies from 19.2\,mag around the Galactic centre to 20.7\,mag over much of the rest of the sky (the median over HEALpixels is 20.5\,mag). 

We apply our inference to all sources in \release\ that have parallaxes. As our prior only reflects single stars in the Galaxy, our distances will be incorrect for the small fraction of extragalactic source in the Gaia catalogue, and may also be wrong for some unresolved binaries, depending on their luminosity ratios.

As some readers may be familiar with our previous catalogue using \gdr{2} data (paper~IV), here is a summary
of the main changes in the new method (which we describe fully in  section~\ref{sec:method}).
\begin{enumerate}
\item We update the source of our prior from a mock catalogue of \gdr{2} \citep{2018PASP..130g4101R} to one of \release\ \citep{2020PASP..132g4501R}.
\item We replace the one-parameter exponential decreasing space
  density (EDSD) distance prior with a more more flexible three-parameter distance prior (section~\ref{sec:distance_prior}).
\item  We again fit the distance prior to a mock catalogue, but 
we no longer use spherical harmonics to smooth the length scale
  of the prior over the sky. We instead adopt a common
  distance prior for all stars within a small area (level 5 HEALpixels).
\item We introduce photogeometric distances (section~\ref{sec:photogeometric_distance}) using a model for the (extincted) colour-absolute magnitude diagram, also defined per HEALpixel (section~\ref{sec:QG_prior}).
\item In paper~IV we summarized each posterior with the mode and the
  highest density interval (HDI). 
  The mode has the disadvantage that it is not invariant
  under nonlinear transformations. This means that 
  if we inferred \rmode\ as the mode of the posterior in distance, then $5\logten\rmode-5$ would not, in general, be the mode of the
  posterior in distance modulus. This is also the case for the mean.
  The quantiles of a distribution, in contrast, are invariant under (monotonic) nonlinear transformations.
   We therefore provide the median (the 50th
  percentile) of the posterior as our distance
  estimate. To characterize the uncertainty in this we quote the 14th and 86th percentiles (an equal-tailed interval, ETI). These are therefore also the quantiles on the absolute magnitude inferred from the distance.
\end{enumerate}

In the next section we describe our method and the construction of the priors.
In section~\ref{sec:results_mock} we apply our method to the \mock\ catalogue, giving some insights into how it performs. We present the results on \release\ in section~\ref{sec:results}, and 
describe the resulting distance catalogue in section~\ref{sec:catalogue} along with its use and limitations. We
summarize in section~\ref{sec:summary}.
Auxiliary information, including additional plots for all HEALpixels, for both the prior and the results, can be found online\footnote{\url{http://www.mpia.de/~calj/gedr3_distances.html}}.

\vspace*{1em}
\section{Method}\label{sec:method}

For each source we compute the following two posterior probability density functions (PDFs) over the distance \dist
\begin{alignat}{4}
\textrm{Geometric:} \ \ && P_{\rm g}^*(\dist \given \parallax, \sigparallax, \hp) && \nonumber \\
\textrm{Photogeometric:} \ \ && P_{\rm pg}^*(\dist \given \parallax, \sigparallax, \hp, \, && \gmag, \cc) \nonumber
\end{alignat}
where \parallax\ is the parallax,
\sigparallax\ is the uncertainty in the parallax,
\hp\ is the HEALpixel number (which depends on Galactic latitude and longitude),
\gmag\ is the apparent magnitude, and
\cc\ is the \bprp\ colour.
The parallax and apparent magnitude will be adjusted to accommodate known issues with the \release\ data, as detailed below. 
The star $^*$ symbol indicates that we infer unnormalized posteriors.
The geometric posterior uses just a distance prior. 
The photogeometric posterior uses this distance prior as well as a colour--magnitude prior that we explain below.
The posteriors are summarized using quantiles computed by Markov Chain Monte Carlo (MCMC) sampling.

\subsection{Geometric distance}

The unnormalized posterior PDF is the product of the likelihood and prior:
\begin{equation}
  P_{\rm g}^*(\dist \given \parallax, \sigparallax, \hp) \,=\  P(\parallax \given \dist, \sigparallax) \, P(\dist \given \hp) \ .
  \label{eqn:distpost_geo}
\end{equation}
The likelihood is conditionally independent of \hp.
We chose to make the second term, which we define in section~\ref{sec:distance_prior}, independent of \sigparallax.

\subsection{Likelihood}\label{sec:likelihood}

Under the assumption of Gaussian parallax uncertainties the likelihood is
\begin{equation}
P(\parallax \given \dist, \sigparallax) \,=\, \frac{1}{\sqrt{2 \pi}\sigparallax} \exp{ \left[ -\frac{1}{2\sigparallax^2}\left(\parallax-\parallaxzp-\frac{1}{r}\right)^2 \right] } 
\label{eqn:likelihood}
\end{equation}
where \parallaxzp\ is the parallax zeropoint. In paper~IV we adopted a constant value of $-0.029$\,mas for this zeropoint, as recommended in the \gdr{2} release. For \release\ the \gaia\ team has published a more sophisticated parallax zeropoint based on analyses of quasars, binary stars, and the Large Magellanic Cloud (LMC) \citep{lindegren_etal_gedr3_parallaxzp}. This is a function of \gmag, the ecliptic latitude, and the effective wavenumber used in the astrometric solution. Ideally this last term was derived from the \bprp\ colour, and this is the case for the standard 5-parameter (5p) astrometric solutions used for 585 million sources \citep{gedr3_release}. But where \bprp\ was unavailable or deemed of insufficient quality, the effective wavenumber was derived as a sixth parameter in the astrometric solution (6p solutions) \citep{lindegren_etal_gedr3_astrometry}, which is the case for 882 million sources. Overall the zeropoint ranges between about $-0.150$ and $+0.130$\,mas (it is narrower for the 5p solutions), although the RMS range is only 0.020\,mas. 
We use this zeropoint correction in equation~\ref{eqn:likelihood}. Our geometric distances are therefore weakly conditioned also on \gmag\ and \cc, but we omit this in the mathematical notation for brevity. For the 2.5 million sources that have parallaxes but no \gmag\ (strictly, no {\tt phot\_g\_mean\_mag}), we use the \release\ global zeropoint of $-0.017$\,mas \citep{lindegren_etal_gedr3_astrometry}.

\subsection{Distance prior}\label{sec:distance_prior}

\begin{figure*}
\begin{center}
\includegraphics[width=0.45\textwidth, angle=0]{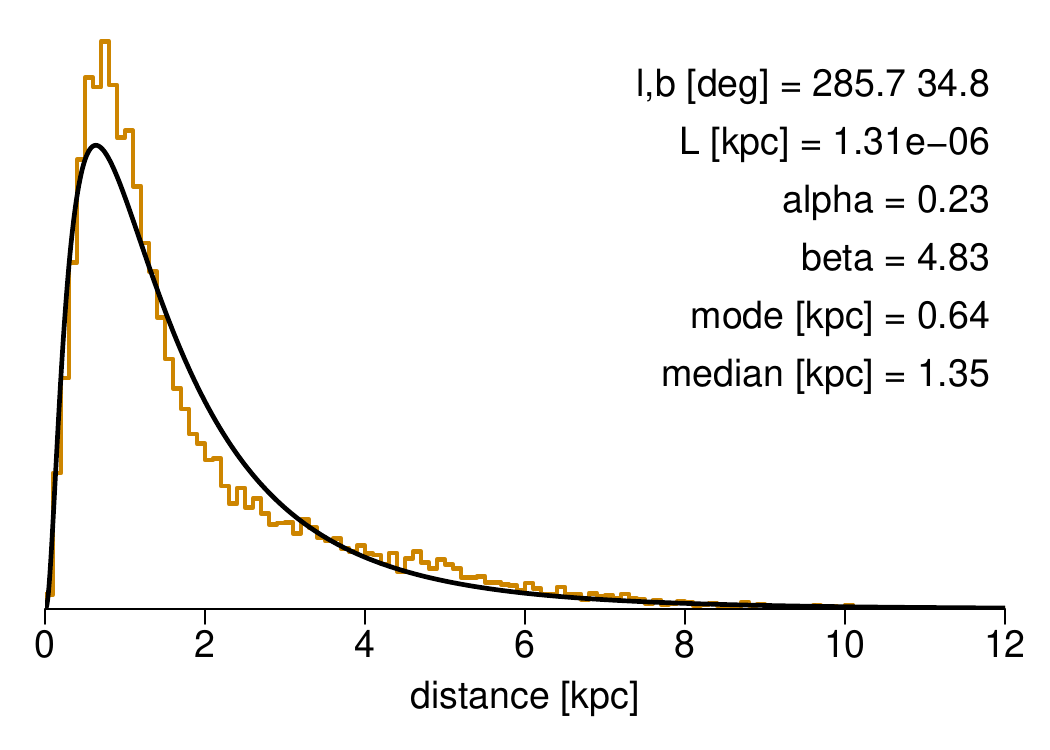}
\includegraphics[width=0.45\textwidth, angle=0]{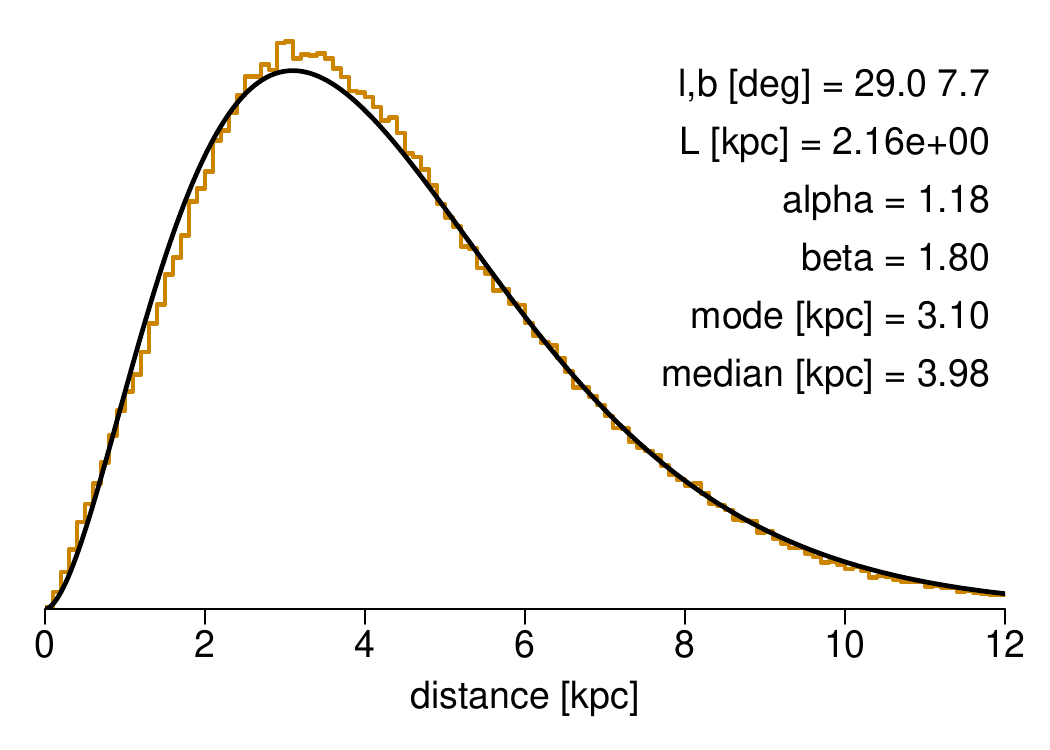}
\caption{Distance priors for two HEALpixels, number 6200 at high latitude (left) and number 7593 at low latitude (right). The histograms show the distributions of the data in the mock catalogue. The smooth curves are the fit of the Generalized Gamma Distribution (GGD; equation~\ref{eqn:GGDprior}) to these data, which defines the distance prior $P(\dist \given \hp)$ with the parameters \rlen, $\alpha$, and $\beta$.
Similar plots for all HEALpixels are available with the auxiliary information online.
\label{fig:GGDprior_fits}}
\end{center}
\end{figure*}

In paper~IV we used the one-parameter EDSD distance prior, which models the space density of stars as dropping exponentially away from the Sun according to a (direction-dependent) length scale. 
Here we adopt the more flexible, three-parameter 
Generalized Gamma Distribution (GGD), which can be written as
\begin{equation}
  P(\dist \given \hp)  \ = \  \begin{dcases}
  \ \frac{1}{\Gamma(\frac{\beta+1}{\alpha})}\frac{\alpha}{\rlen^{\beta+1}}\,\dist^\beta\,e^{-(\dist/\rlen)^\alpha}  & \:{\rm if}~~ \dist \geq 0 \\
   \ 0                          & \:{\rm otherwise}
\end{dcases}
\label{eqn:GGDprior}
\end{equation}
for $\alpha > 0$, $\beta > -1$, and $\rlen > 0$. $\Gamma()$ is the gamma function.
This PDF is unimodal with an exponentially decreasing tail to larger distances. 
The mode is $\rlen(\beta/\alpha)^{1/\alpha}$ for $\beta>0$, and zero otherwise.
The EDSD is a special case of the GGD with $\alpha=1, \beta=2$.
We fit the GGD prior for each HEALpixel separately via maximum likelihood using stars from the mock catalogue. The HEALpixel (\hp) dependency on the left side of equation~\ref{eqn:GGDprior} is equivalent to a dependency on $\alpha, \beta, \rlen$.

\begin{figure}
\begin{center}
\includegraphics[width=0.50\textwidth, angle=0]{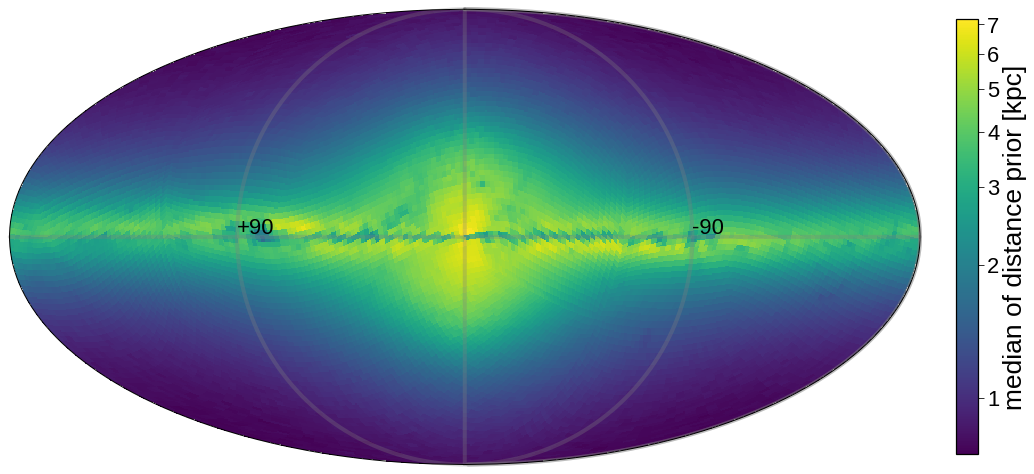}
\caption{The variation of the median of the distance prior over the sky shown in Galactic coordinates on a Mollweide equal-area projection. The LMC/SMC are excluded from our prior.
\label{fig:GGDprior_median_sky}}
\end{center}
\end{figure}

Example fits for two HEALpixels, one at low Galactic latitude and one at high Galactic latitude, are shown in Figure~\ref{fig:GGDprior_fits}.  Although the GGD prior provides a better fit than the EDSD prior -- which is why we use it -- the parameter \rlen\ may no longer be interpreted as a meaningful length scale, because it varies from 
3e-7 to 1e4\,pc over all HEALpixels. The appropriate characteristic scale of the GGD prior in this work is its median, for which which there is no closed-form expression.
The median varies between 745 and 7185\,pc depending on HEALpixel (Figure~\ref{fig:GGDprior_median_sky}). Fits for each HEALpixel can be found in the auxiliary information online.

In the limit of uninformative parallaxes, the geometric posterior converges on the GGD prior, and so the median distance converges on the median of this prior. In paper IV this convergence was on the mode of the EDSD prior.  For the prior fits used in the present paper, the ratio of the GGD median to the EDSD mode ranges from 1.17 to 1.57. There are potential improvements one could make to the prior to give a better convergence in the limit of poor data. Some considerations are in appendix~\ref{sec:thoughts_prior}.

\subsection{Photogeometric distance}\label{sec:photogeometric_distance}

\begin{figure*}
\begin{center}
\includegraphics[width=0.49\textwidth, angle=0]{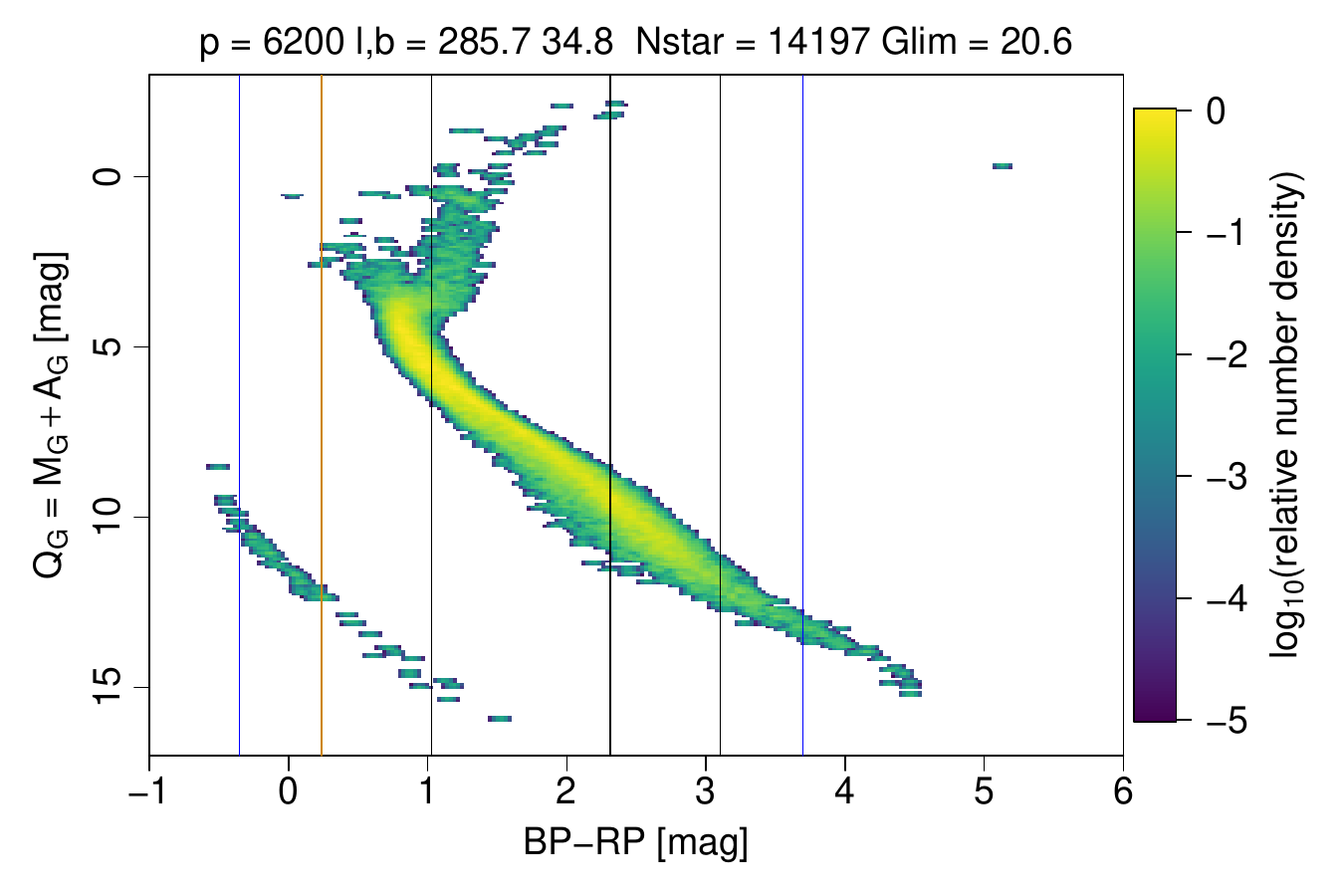}
\includegraphics[width=0.49\textwidth, angle=0]{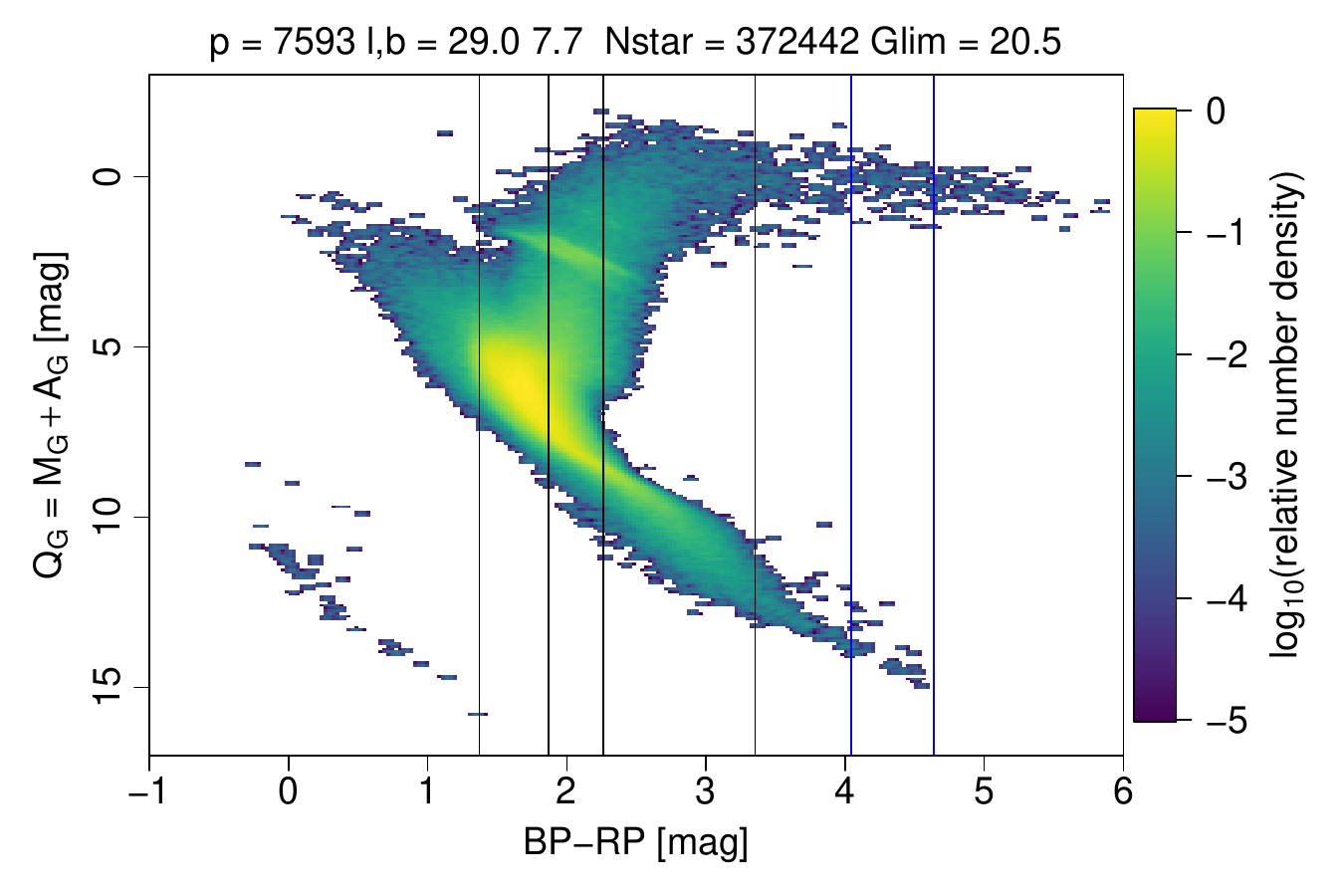}
\caption{CQDs for HEALpixels 6200 (left) and 7593 (right) in the mock catalogue. The density of stars is shown on a logarithmic colour scale relative to the maximum density in each HEALpixel (so the zero point of the density scales are not the same in the two panels). The text at the top of each panel gives the Galactic longitude and latitude ($\glon, \glat$) of the centre of the HEALpixel in degrees, the number of stars, and the faintest magnitude. The vertical lines identify particular \QG-models that are shown in Figures~\ref{fig:QGmodels1} and~\ref{fig:QGmodels2}.
Similar plots for all HEALpixels are available with the auxiliary information online.
\label{fig:mockCQD}}
\end{center}
\end{figure*}

We define the quantity \QG\ as
\begin{equation}
  \QG \equiv \MG + \AG \,=\, \gmag - 5\logten\dist + 5 \ .
 \label{eqn:continuity}
\end{equation}
The equality ($=$), which is a statement of flux conservation, holds only when all the quantities are noise-free.
If we knew \QG\ for a star, then a measurement of \gmag\ gives us an estimate of \dist. Given that the uncertainties on \gmag\ in \release\ are generally less than a few millimagnitudes
(0.3 to 6\,mmag for $\gmag < 20$\,mag; \citealt{gedr3_release}),
this would be a reasonably precise estimate. We do not know \QG, but we can take advantage of the fact that the two-dimensional colour--\QG\ space for stars is not uniformly populated.
This space (e.g.\ Figure~\ref{fig:mockCQD}) which we call the CQD -- in analogy to the CMD (colour--magnitude diagram) -- would be identical to the colour-absolute-magnitude diagram if there were no interstellar extinction. Thus if we know the \bprp\ colour of the star, this diagram places limits on possible values of \QG, and therefore on the distance to the star. We will use the mock catalogue to model the CQD (per HEALpixel) and from this compute a prior over \QG\ given the magnitude and colour of the star.

The formal procedure is as follows, initially making no assumptions about \gmag. We assume the colour to be effectively noise-free. This is reasonable given the relatively low noise for most sources (13 to 120\,mmag for $\gmag < 20$\,mag; \citealt{gedr3_release}), and the fact that the prior is anyway imperfect (see section~\ref{sec:QG_prior}).
Using Bayes' theorem, the unnormalized posterior we want to estimate can be decomposed into a product of two terms
\begin{equation}
  P_{\rm pg}^*(\dist \given \parallax, \sigparallax, \gmag, \cc, \hp) \,=\  P(\parallax \given \dist, \sigparallax) \, P(\dist \given \gmag, \cc, \hp) \ .
  \label{eqn:distpost_unnormalized}
\end{equation}
The first term on the right side is the parallax likelihood (section~\ref{sec:likelihood}). It is independent of \gmag, \cc, and \hp\ once it is conditioned on \sigparallax, which is estimated in the \gaia\ astrometric solution using quantities that depend on the magnitude, colour, scanning law, etc.\ \citep{lindegren_etal_gedr3_astrometry}. The second term is independent of the parallax measurement process and thus of \parallax\ and \sigparallax. We may write this second term as a marginalization over \QG\ and then apply Bayes' theorem as follows
\begin{alignat}{2}
  & P(\dist \given \gmag, \cc, \hp)  \label{eqn:distpost_part} \\
                                            \,&=\  \int P(\dist, \QG \given \gmag, \cc, \hp) \ \deriv\QG \nonumber \\
                                            \,&=\  \int \frac{1}{P(\gmag \given \cc, \hp)}  \, P(\gmag \given \dist, \QG) \, P(\dist, \QG \given \cc, \hp) \ \deriv\QG \nonumber \\
                                            \,&=\  \frac{P(\dist \given \cc, \hp)}{P(\gmag \given \cc, \hp)}  \int P(\gmag \given \dist, \QG) \, P(\QG \given \dist, \cc, \hp) \ \deriv\QG \ . \nonumber 
\end{alignat}
In the last line, the first term under the integral is formally the likelihood for \gmag\ (and is conditionally independent of \cc\ and \hp\ due to equation~\ref{eqn:continuity}). However, as \gmag\ is measured much more precisely than the intrinsic spread in \QG\ -- that is, the second term under the integral is a much broader function -- we can consider \gmag\ to be noise-free to a good approximation. This makes the first term a delta function and so 
the integral is non-zero only when equation \ref{eqn:continuity} is satisfied.

We make two further assumptions about the terms in the last line of equation~\ref{eqn:distpost_part}. The first is to make the distance prior independent of colour, i.e.\ $P(\dist \given \cc, \hp) \rightarrow P(\dist \given \hp)$. This is now the same distance prior as used in the geometric posterior (equation~\ref{eqn:distpost_geo}). The second is to assume that the CQD is independent of distance, i.e.\ $P(\QG \given \dist, \cc, \hp)\rightarrow P(\QG \given \cc, \hp)$. This is not true in general, but we chose not to add this extra layer of dependence on \gedr3mock\ (see section~\ref{sec:QG_prior}).

With these assumptions, the (unnormalized) posterior in equation~\ref{eqn:distpost_unnormalized} can now be written as
\begin{alignat}{2}
 & P_{\rm pg}^*(\dist \given \parallax, \sigparallax, \gmag, \cc, \hp) \, \simeq\     
 P(\parallax \given \dist, \sigparallax) \, P(\dist \given \hp) \, \times  \nonumber \\ 
  & \hspace*{7em} P(\QG = \gmag - 5\logten\dist + 5 \given \cc, \hp) \ . \label{eqn:distpost_unnormalized_2} 
    \end{alignat}
The missing normalization constant, $1/P(\parallax, \gmag \given \cc, \hp)$, is not required.
This posterior is simply the geometric posterior (equation~\ref{eqn:distpost_geo}) multiplied by an additional prior\footnote{We could have arrived at this without using the marginalization in equation~\ref{eqn:distpost_part} if we assumed \gmag\ to be noise-free from the outset. But the marginalization justifies how small the noise in \gmag\ has to be for this to be valid.} over \QG.

\subsection{\QG\ prior}\label{sec:QG_prior}

\begin{figure*}
\begin{center}
\includegraphics[width=1.00\textwidth, angle=0]{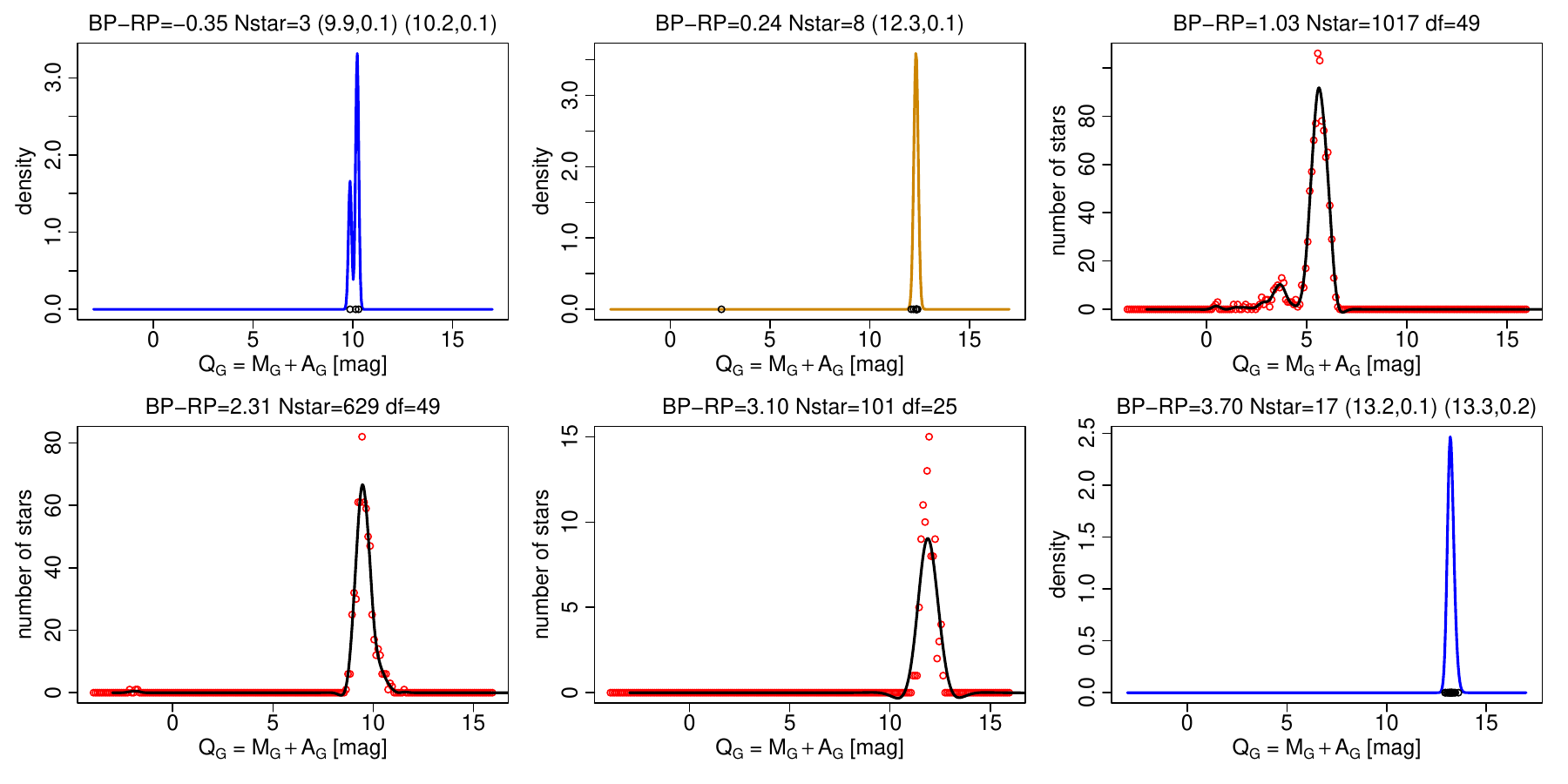}
\caption{\QG\ prior models constructed from the CQD of HEALpixel 6200. Each of the six panels shows a fit to the mock data at a different \bprp\ colour, corresponding to the six vertical stripes shown in Figure~\ref{fig:mockCQD} (left panel). Model fits using smoothing splines are plotted as black lines with the degrees of freedom (df) as indicated and the (binned) data in the fit show as red circles. Model fits using one or two Gaussian components are plotted as orange and blue lines respectively, with the data in the fit shown as black circles and the mean and standard deviation of the fit components indicated in parentheses at the top of each panel. These density functions show the prior PDF $P(\QG \given \cc, \hp)$ at discrete colours before imposing the minimum threshold which ensures the prior density is always greater than zero. Similar plots for all colour strips in HEALpixels are available with the auxiliary information online.
\label{fig:QGmodels1}}
\end{center}
\end{figure*}

\begin{figure*}
\begin{center}
\includegraphics[width=1.00\textwidth, angle=0]{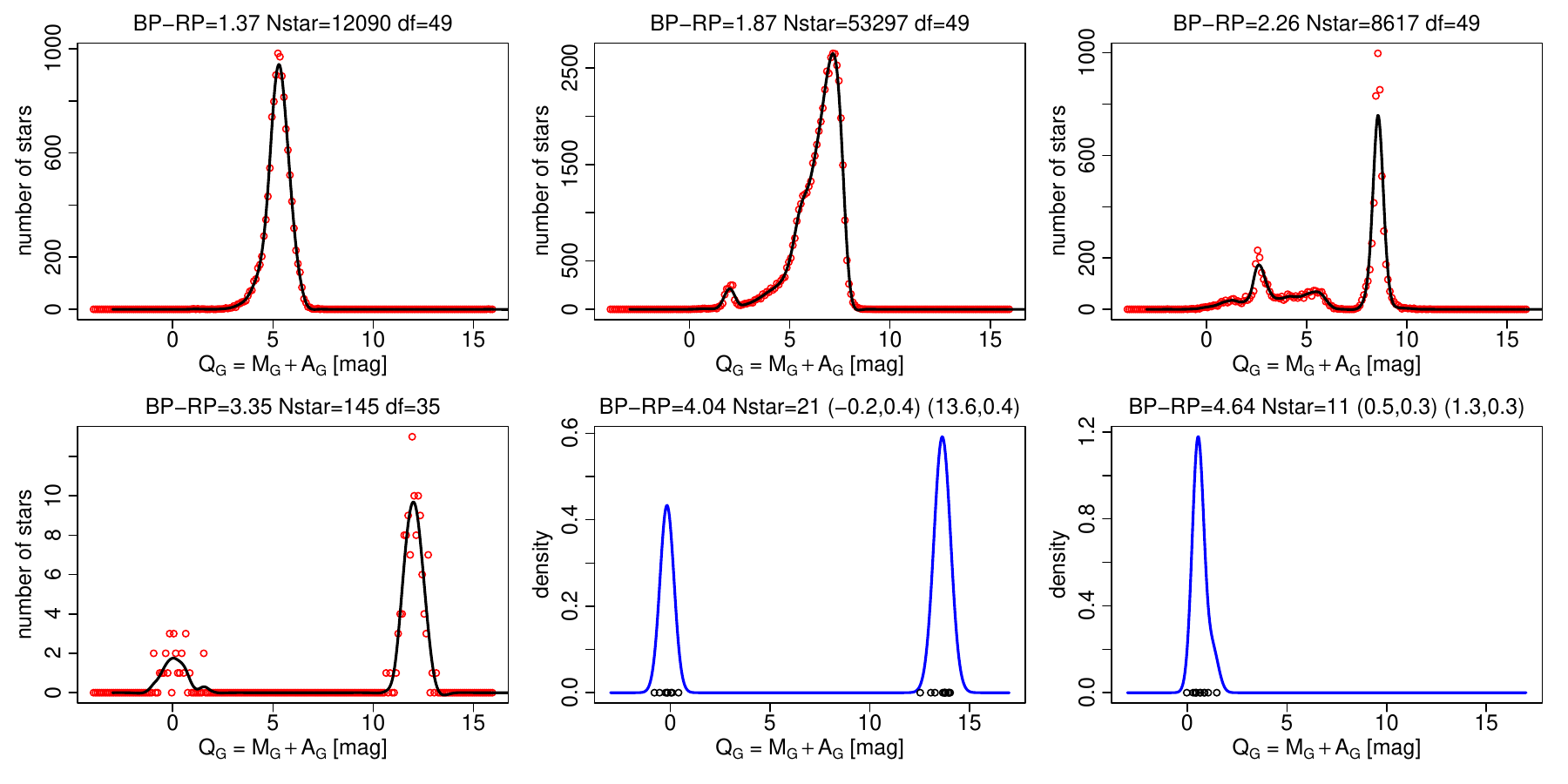}
\caption{As Figure~\ref{fig:QGmodels1} but now for HEALpixel 7593, the CQD of which is in the right panel of Figure~\ref{fig:mockCQD}.
\label{fig:QGmodels2}}
\end{center}
\end{figure*}

We construct the prior P(\QG \given \cc, \hp) from the mock catalogue. Given the complexity of the CQD and its variation over the sky, we do not attempt to fit the prior as a continuous 3D (position and colour) parametric function.
We instead compute a CQD for each HEALpixel, two examples of which are shown in Figure~\ref{fig:mockCQD}.
Within each we compute a series of one-dimensional functions at a series of colours in the following way. We divide the full colour range of a given HEALpixel into strips of 0.1\,mag width in colour, then for each strip fit a model to the stellar number density as a function of \QG\ (now ignoring the colour variation in each strip). If there are more than 40 stars in a strip, we bin the data into bins of 0.1\,mag and fit a smoothing spline with 
$\min(\lfloor{N/4}\rfloor, 50)$ degrees of freedom (df), where $N$ is the number of stars in the strip (which can be many thousands).
If there are fewer than 40 stars we cannot fit a good spline. 
This generally occurs at the bluest and reddest ends of the CQD. Here the \QG\ distribution is often characterized by two widely-separated components, either the main sequence (MS) and white dwarf (WD) branches, or the MS and giant star branches (see Figure~\ref{fig:mockCQD}).
Thus when $N<40$  we instead fit a two-component Gaussian mixture model, with the constraint that the minimum and maximum standard deviation of each component be $\sigma_{\rm min}=0.08$\,mag and $\sigma_{\rm max}=1.0$\,mag respectively.
A full fit requires at least five stars, so if there are as few as two stars we constrain the solution to first have equal standard deviations and then to have standard deviations of $\sigma_{\rm min}$. If $N=1$ our model is a one-component Gaussian with mean equal to the \QG\ of the star and standard deviation equal to $\sigma_{\rm min}$. If there are no stars the model is null. 
Examples of the fits are shown in Figures~\ref{fig:QGmodels1} and~\ref{fig:QGmodels2}.
 
As a smoothing spline can give a negative fit, and both these and the Gaussian models can yield very small values for the density, we impose that the minimum density is never less than $10^{-3}$ of the integrated density (computed prior to fitting the model). Thus our prior is nowhere zero, meaning that even if the data indicate a \QG\ in the regions where the mock catalogue is empty, the posterior will not be zero. 
This allows sources to achieve distances that place them outside the occupied regions of the mock CQD.

For a given HEALpixel, each prior model refers to a specific colour, namely the centre of a 0.1\,mag-wide strip.
This is larger than the uncertainty in the colour for all but the faintest \release\ sources.
When evaluating the prior during the inference process, we compute \QG\ from equation~\ref{eqn:continuity}, evaluate the densities of the two priors that bracket its colour, then linearly interpolate. This ensures that our prior is continuous in colour. If one of the models is null we use the other model as is. If both models are null, or if the source is outside of the colour range of the mock CQD, we do not infer a photogeometric distance.
The flag field in our catalogue indicates what kind of \QG\ models were used (see section~\ref{sec:catalogue}).

The computation of \QG\ in equation~\ref{eqn:continuity} requires the G-band magnitude of the source. For this we use the {\tt phot\_g\_mean\_mag} field in \release\ corrected for the processing error described in section 8.3 of \cite{riello_etal_gedr3_photometry}. This correction, which is a function of magnitude and colour, can be as large as 25\,mmag. 

\subsection{Posterior sampling and summary}\label{sec:sampling}

The posteriors are formally the answer to our inference process.
The geometric posterior has a simple parametric form which may be computed by the reader using the data in the \release\ catalogue and the parameters of our prior (available with the auxiliary information online).
The photogeometric posterior is generally non-parametric.
Both posteriors are asymmetric and not necessarily unimodal (section~\ref{sec:multimodality}).

There are a variety of statistics one could use to summarize these PDFs, such as the mean, median, or mode.
There is no theoretically correct measure, and all have their drawbacks. 
We use quantiles, primarily because they are invariant under nonlinear transformations, and 
so are simultaneously the quantiles of the posterior in distance modulus, $5\logten\dist-5$.
We use the three quantiles at 0.159, 0.5, and 0.841, 
which we label  \rlo, \rmed, and \rhi\ respectively. The central quantile is the median. The outer two quantiles give a 68\% confidence interval around the median. The difference between each quantile and the median is a Gaussian 1$\sigma$-like estimate of the uncertainty. Due to the intrinsic asymmetry of the posteriors we report the lower and upper values separately.

\subsubsection{Markov Chain Monte Carlo}\label{sec:MCMC}

Neither the geometric nor photogeometric posteriors have closed-form expressions for their quantiles so we must compute these numerically.
We do this using Markov Chain Monte Carlo (MCMC), specifically the Metropolis algorithm. 

We adopt the following scheme for the MCMC initialization and step size.
We first compute the geometric distance posterior using the EDSD prior from paper IV. The length scale of this prior is set to 0.374\rmed, where \rmed\ is the median distance of the stars in the mock catalogue for that HEALpixel.\footnote{In paper IV we used $(1/3)\rmed$, as the maximum likelihood fit of the length scale is a third of the mean. However, the median is a slightly biased estimator of the mean for the EDSD. For the typical length scales involved we found empirically that the mean is about 12\% (0.374/0.333) larger than the median.} We use the mode of this posterior, \rEDSDmode, which has a closed-form solution (paper I), as the initialization for the geometric posterior. 
The initialization scheme for the photogeometric posterior is more complicated, in accordance with its more complicated shape, and depends on \rEDSDmode, fractional parallax uncertainty (fpu, \fpu), and the characteristic length scale of the \QG\ prior model(s). 

For both types of posterior the step size needs to be adapted to the characteristic width of the posterior, which is generally wider the larger the fpu.
We found a suitable step size to be
$(3/4)\rinit\times \min(|\sigparallax/\parallax|, 1/3)$, where \rinit\ is the initialization value.

This scheme allows relatively short burn-ins: we use just 50. We experimented with chains of various chain lengths, employing various tests of convergence. Longer chains are always better, but as we need to sample around three billion posteriors, some parsimony is called for.  We settled on 500 samples (post burn-in).  Although the chains are not always settled, they are generally good enough to compute the required quantiles with reasonable precision. To quantify this we obtained 20 different MCMC chains 
and computed the standard deviation of the median distance estimates and half the mean of the confidence intervals. The ratio of these is a measure of the convergence noise. Doing this for thousands of stars we find this to be between 0.1 and 0.2 in general. For the geometric posteriors in particular it can be larger for fractional parallax uncertainties larger than 0.3.


\subsubsection{Multimodality}\label{sec:multimodality}

The posteriors can be multimodal.
This is more likely to be the case for the photogeometric posterior at large fpu, as its prior can be multimodal. Multimodality is very rare for the geometric posterior.

Although multimodality is a challenge for MCMC sampling methods, we find that even widely-separated modes can be sampled in our scheme. Our 68\% confidence interval often encompasses the span of such multimodality. This is a blessing and a curse: the distance precision in a single mode may be quite good, yet a large confidence interval is obtained due to the presence of a second mode.
To assist in identifying possible multimodality we perform the Hartigan dip test \citep{hartigan1985}. This is a classical statistical test in which the null hypothesis is a unimodal posterior, i.e.\ a small p-value suggests the distribution may not be unimodal. We select a threshold of $10^{-3}$ and set a flag to 1 if the p-value is lower than this, thereby suggesting possible multimodality.
If the p-value is above this threshold or the test does not work for any reason, the flag is 0.
The test is not particularly accurate and should not be over-interpreted.
Furthermore, it is done on the MCMC samples, not on the true posterior, so tends to be raised more often than expected due to the intrinsic noise of MCMC sampling.

\vspace*{1em}
\section{Performance on the mock catalogue}\label{sec:results_mock}

Before looking at the results on \release, we evaluate the performance of our method using the mock catalogue, as here we know the true distances. In doing this we add Gaussian random noise to the parallaxes using the {\tt parallax\_error} field in \mock, which is a model of the expected uncertainties in the \release\ parallaxes.  
As the data are drawn from the same distance distribution and CQD from which the prior was constructed, this is a somewhat optimistic test, despite the noise.
Unless noted otherwise, throughout this section the term ``fpu" refers to the {\em true} fractional parallax uncertainty, i.e.\ that computed using the true parallax

\subsection{Example posteriors}

\begin{figure*}
\begin{center}
\includegraphics[width=1.0\textwidth, angle=0]{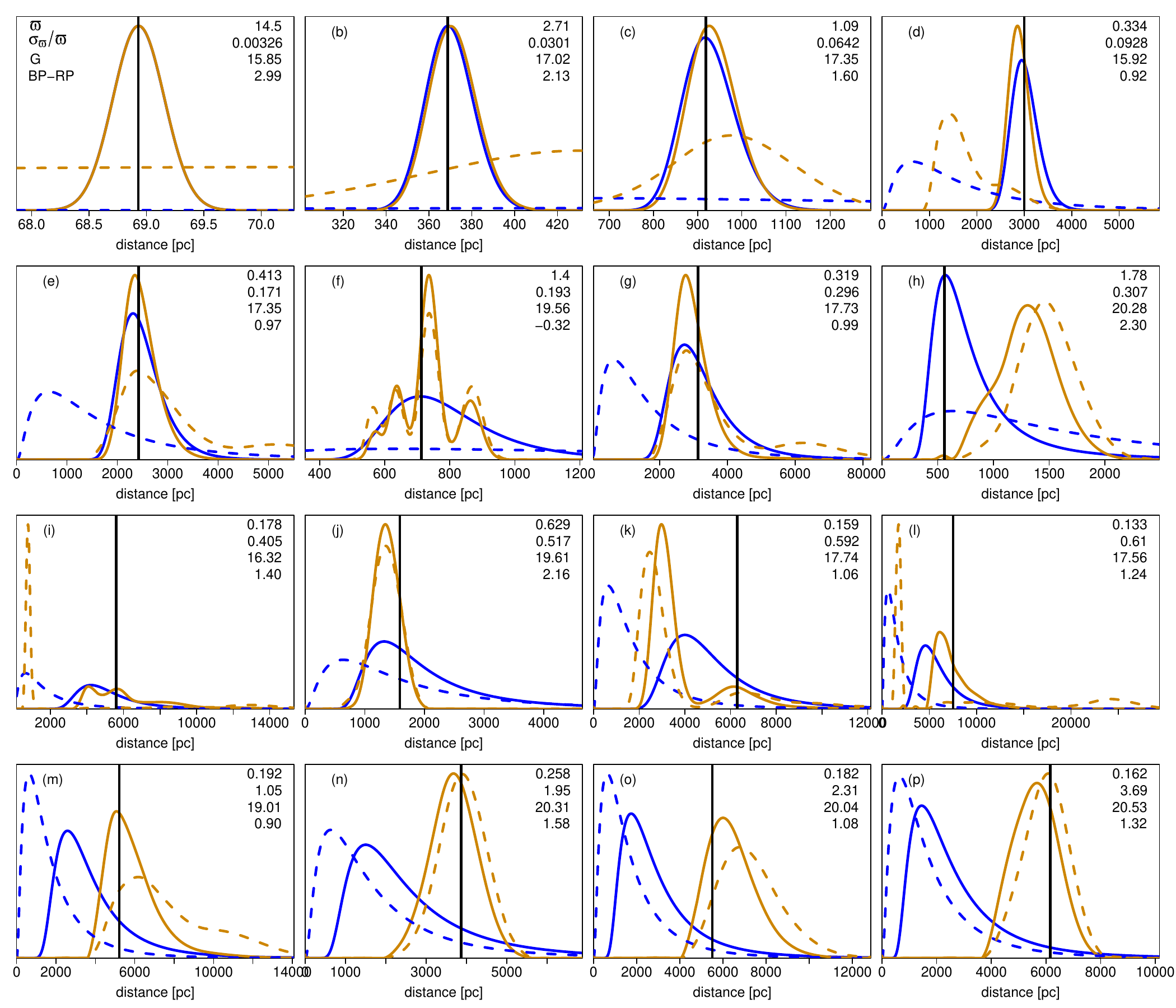}
\caption{Example normalized posteriors (solid lines) and corresponding normalized priors (dashed lines) for geometric distances (blue) and photogeometric distances (orange) for various stars in the mock catalogue (one per panel). These have been selected to show the variety; they are not a random subset. The vertical solid line is the true distance. The inverse of this is not the parallax seen by the inference, because noise was added. All stars are from HEALpixel 6200, so the distance prior (blue dashed line) is the same in all panels. 
The four numbers in the top-right corner of each panel are, from top to bottom, \parallax, true fpu, \gmag, and \bprp.
Stars are ordered by increasing fpu. The two posteriors coincide in the top-left panel.
\label{fig:mock_inference_posteriors}
}
\end{center}
\end{figure*}

Figure~\ref{fig:mock_inference_posteriors} shows examples of both types of posterior compared to their priors. At small fpu, e.g.\ panels (a) to (c), the two posteriors are very similar, with a median (and mode) near to the true distance, shown as the vertical line. As long as the fpu is not too large, the prior plays little role and the posterior can be quite different, e.g.\ panel (d), although this can also occur at larger fpu, e.g.\ panels (i) and (l). Panel (f) shows a multimodal photogeometric prior and posterior. The two types of prior sometimes disagree, as can the posteriors. In panel (h), which is for a 30\% parallax uncertainty, the geometric posterior is more consistent with the true distance. Note that the parallax that the algorithm sees does not correspond to the vertical line, so for large fpu we cannot expect either posterior to peak near this. Panel (k) shows a multimodal posterior in which the true distance is close to a smaller mode. This happens here because the parallax has 50\% noise, so the measured parallax corresponds to a smaller distance (where both geometric and photogeometric posteriors peak). At larger fpu -- the bottom row is all for more than 1.0 -- the photogeometric prior is often more consistent with the true distance than the geometric one.

\subsection{Comparison to truth}

\subsubsection{Qualitative analysis}\label{sec:mock_qualitative_analysis}

\begin{figure*}[p]
\begin{center}
\includegraphics[width=1.00\textwidth, angle=0]{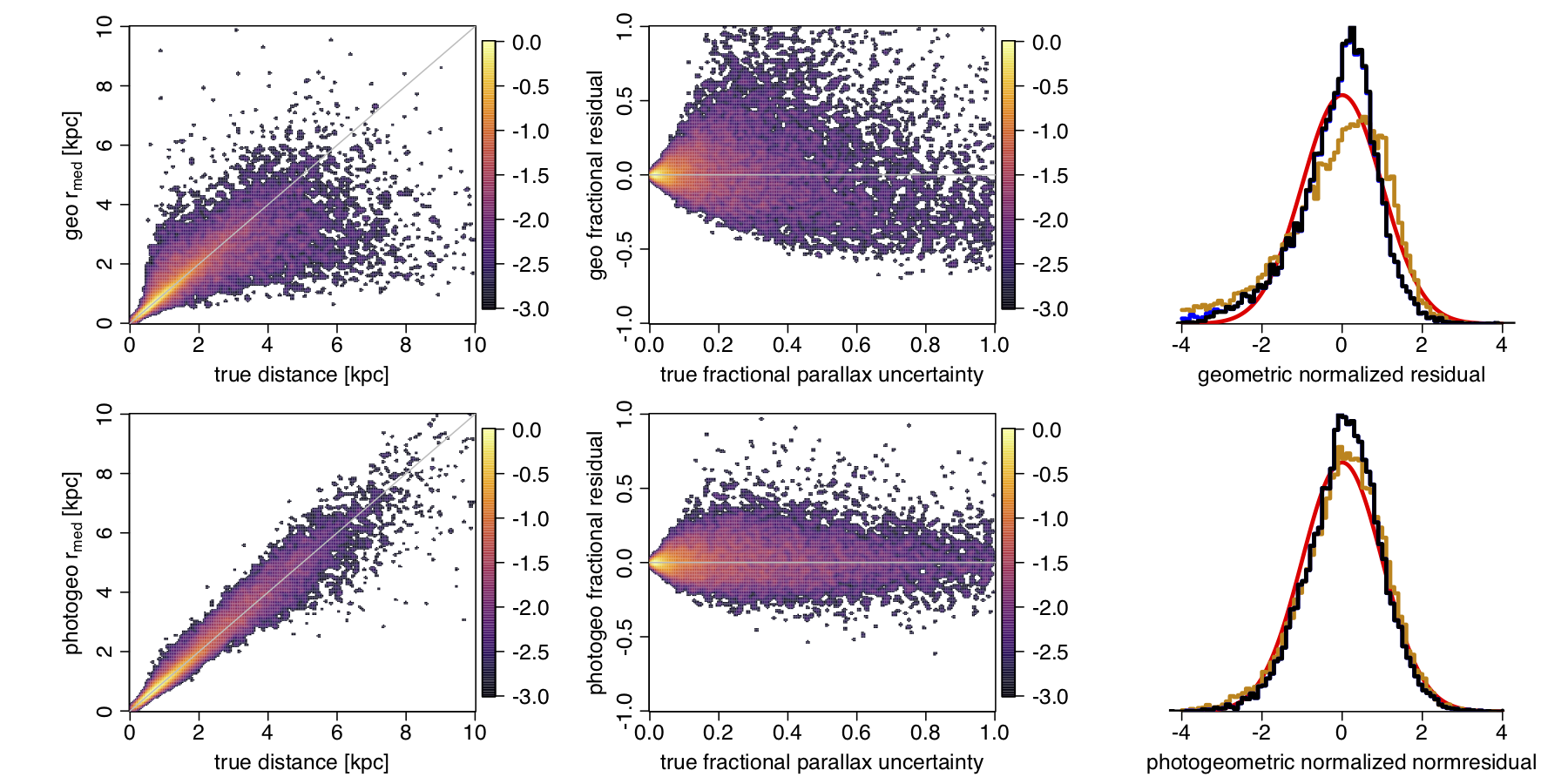}
\caption{Results of the distance inference on mock catalogue HEALpixel 6200. The top row shows geometric distances, the bottom row photogeometric ones. The left column compares the inferred distances (vertical axis) to the true distances for all sources. This cover the full range of fractional parallax uncertainties, which has a median of 0.20 and central 90\% range of 0.03--1.08.  The middle column shows the fractional distance residuals as a function of the true fractional parallax uncertainty (fpu). 
In these first four panels the colour scale is a logarithmic density (base 10) scale relative to the highest density cell in each panel.
The right column shows the normalized residuals: the difference between the inferred and the true value, divided by an uncertainty measure. The three colours refer to three uncertainty measures: orange is $\rmed-\rlo$, blue is $\rhi-\rmed$, black is $(1/2)(\rhi-\rlo)$. The blue and black lines virtually coincide. The smooth red curve is a unit Gaussian for comparison.
\label{fig:mock_inference_1}}
\end{center}
\end{figure*}

\begin{figure*}
\begin{center}
\includegraphics[width=1.00\textwidth, angle=0]{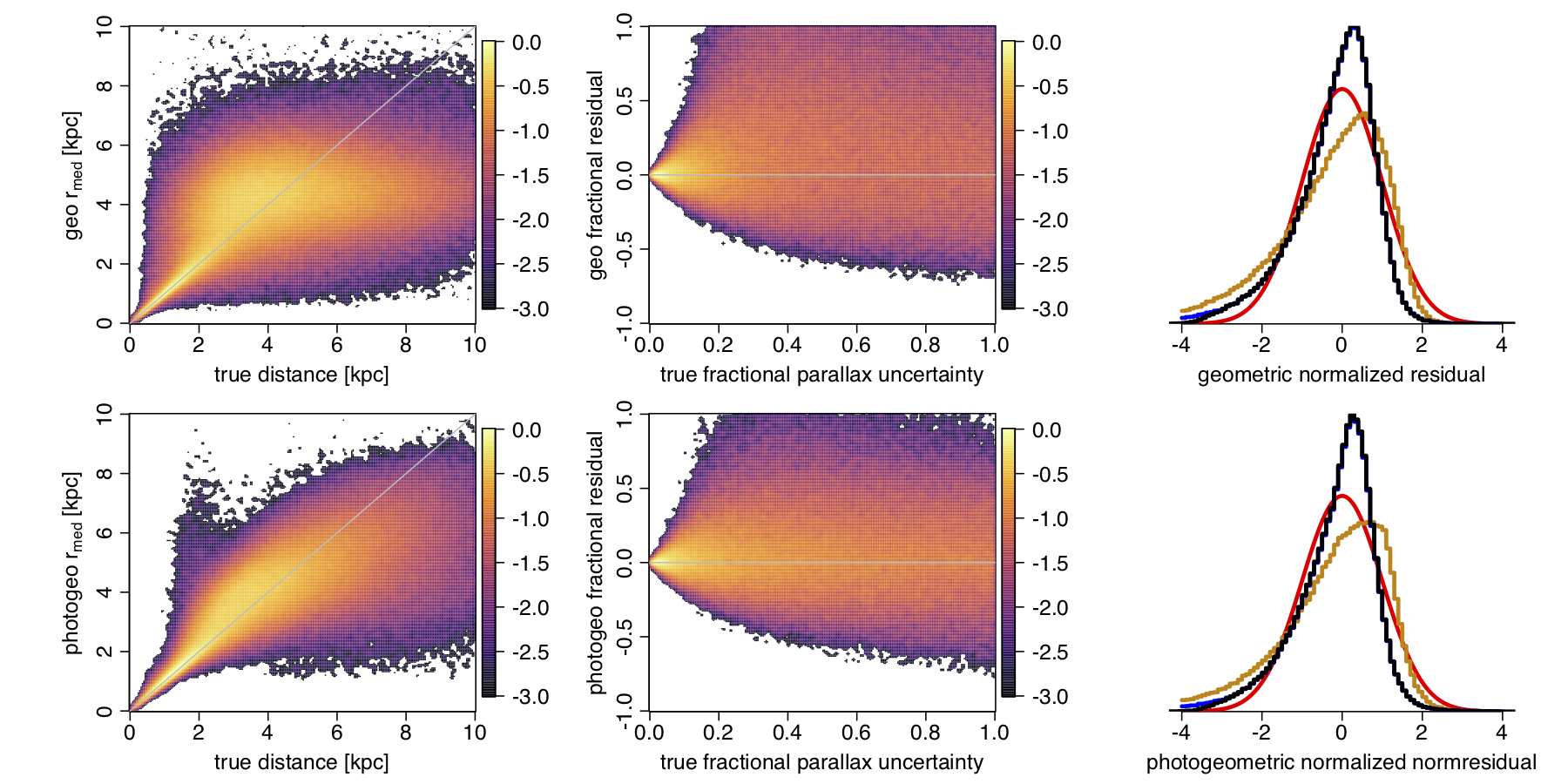}
\caption{As Figure~\ref{fig:mock_inference_1} but now for HEALpixel 7593.
The median fpu is 1.18 and the 90\% range is 0.21--3.57.
\label{fig:mock_inference_2}}
\end{center}
\end{figure*}

Distance inference results for two HEALpixels are shown in Figures~\ref{fig:mock_inference_1} and~\ref{fig:mock_inference_2}. 
We see a good correlation between the inferred and true distances out to several kpc (left columns). The degradation at larger distances is mostly due to stars with larger fpu, as can be seen in the middle columns of these figures. 
The fractional residual is defined as the estimated minus true distance, divided by the true distance.
Note that these middle columns show the {\em true} fpu, i.e.\ as computed from the noise-free parallax, which is not the same as the {\em measured} (noisy) fpu that the inference algorithm encounters. (See section~\ref{sec:poor_parallax_limit} for a consequence of this difference.)
At large fpu the photogeometric distances perform  better than the geometric ones, because even when the parallax is of limited use there is still distance information from the colour and magnitude via the \QG\ model. 
For geometric distances, in contrast, as the measured fpu increases, the distance prior dominates the likelihood, so the median of the posterior is pushed towards the median of the prior.
Hence at large fpu, the geometric distances to stars that are truly more distant than the median of the prior will generally be underestimated.
Faraway stars tend to have larger fpu than nearby stars, because they have both smaller parallaxes and larger parallax uncertainties (as they are fainter). 
Thus as a whole, any {\em underestimation} of geometric distances to stars that are {\em beyond} the median of the prior will tend to be larger than the {\em overestimation} of the geometric distances to stars that are {\em closer} than the median of the prior. This explains why the distribution in the top-left panels of Figures~\ref{fig:mock_inference_1} and~\ref{fig:mock_inference_2} flatten at larger distances.
This feature is suppressed in the photogeometric distances (bottom-left panels) because for large fpu, the \QG\ prior can overrule the geometric prior. We also see more flattening for the low latitude HEALpixel in Figure~\ref{fig:mock_inference_2} than the high latitude HEALpixel in Figure~\ref{fig:mock_inference_1} because the low latitude HEALpixel has larger fpus on average.

The right columns of Figures~\ref{fig:mock_inference_1} and~\ref{fig:mock_inference_2} assess how well the estimated distance uncertainties explain the residuals, by plotting the distribution of residual$/$uncertainty. This is shown using three different representations of the uncertainty. 
The upper uncertainty, $\rhi-\rmed$, and symmetrized uncertainty$(\rhi-\rlo)/2$, shown in blue and black respectively, yield almost identical distributions.
For the high latitude HEALpixel 6200 (Figure~\ref{fig:mock_inference_1}) they are quite close to a unit Gaussian, in particular for the photogeometric estimates. The lower uncertainty,  $\rmed-\rlo$, shown in orange, is negatively skewed (larger tail to negative values),  suggesting that the lower uncertainty measure, \rlo, is slightly underestimated. 
This is more noticeable in the low latitude HEALpixel 7593 (Figure~\ref{fig:mock_inference_2}), where we also see that the photogeometric estimates are slightly more skewed than the geometric ones.

\subsubsection{Quantitative analysis}\label{sec:mock_quantitative_analysis}

To quantify the accuracy of our results we use the median of the fractional distance residual, which we call the {\em bias}, and the median absolute of the fractional distance residual, which we call the {\em scatter}. 
These are robust versions of the mean and standard deviation, respectively. For normally-distributed residuals the mean equals the median, and the standard deviation is 1.48 times the median absolute deviation.

For HEALpixel 6200 the bias and scatter for the geometric distances over all stars are +0.29e-3 and 0.10 respectively. If we limit the computation of these metrics to the 50\% of stars in this HEALpixel with $0 < \fpu < 0.20$, the bias is +5.3e-3 and the scatter is 0.037. The scatter in this subsample is smaller, as expected. The bias is larger because stars with small fpu tend to be nearer stars, whereas the distance prior is characteristic of all the stars, which are more distant on average. Hence the prior pulls up the distances for the small fpu subsample, leading to a more positive bias.

For the photogeometric distances, the bias and scatter over all stars are +5.7e-3 and 0.059 respectively, and for the $0 < \fpu < 0.20$ subsample are +2.5e-3 and 0.032 respectively. 
The scatter over the full sample is smaller for the photogeometric estimates than for the geometric ones, because the former benefit from the additional information in the stars' colours and magnitudes. The situation is particularly fortuitous here because of the near-perfect match between the \QG\ models and the actual distribution of \QG\ in the data. 
For the full sample the bias is larger for the photogeometric distances than for the geometric ones, although still small on an absolute scale.
For the small fpu subsample the photogeometric distances are not much more accurate than the geometric ones, because the parallax dominates the distance estimate.
 
Turning now to the low latitude HEALpixel 7593 (Figure~\ref{fig:mock_inference_2}), the bias and scatter in the geometric distances over all stars are $-0.16$e-3 and 0.27 respectively. There are two reasons for the larger scatter in this HEALpixel. 
The first is that the parallax uncertainties are larger: the median
parallax uncertainty is 0.32\,mas, as opposed to 0.15\,mas in HEALpixel 6200.  This in turn is because the stars are on average 0.9 magnitude fainter in 
HEALpixel 7593 (one reason for which is the larger extinction, as is apparent from Figure~\ref{fig:mockCQD}). 
The second reason is that the median true distance to stars is larger in this low latitude HEALpixel than in the high latitude one (4.0\,kpc vs 1.2\,kpc; see Figure~\ref{fig:GGDprior_fits}). This may seem counter-intuitive, but is a consequence of distant disk (and bulge) stars at low latitudes that remain visible to larger distances despite the higher average extinction. At higher latitudes, in contrast, there are no distant disk stars, and hardly any halo stars (which are scarce in \gaia\ anyway).
Both of these facts contribute to the larger fpu in the low latitude pixel -- median of 1.18, central 90\% range of 0.21--3.57 -- than in the high latitude HEALpixel -- median of 0.20, central 90\% range of 0.03--1.08.
Even if we look at just the 9\% of stars in the low latitude HEALpixel with $0 < \fpu < 0.20$, we get a bias and scatter of +25e-3 and 0.069 respectively, which are still significantly worse than the higher latitude HEALpixel for the same fpu range.

Concerning the photogeometric distances in HEALpixel 7593, the bias and scatter for all stars are $-3.8$e-3 and 0.17 respectively, and for the $0 < \fpu < 0.20$ subsample are +20e-3 and 0.062 respectively. For the full sample we again see a significant decrease in the scatter compared to the geometric distances. In a real application we may get less benefit from the \QG\ prior at low latitudes because our model CQD may differ from the true (unknown) CQD more than at high latitudes, on account of the increased complexity of the stellar populations and interstellar extinction near the Galactic plane.

\subsection{Inferred CQDs}\label{sec:mock_inferered_CQDs}

\begin{figure*}[p]
\begin{center}
\includegraphics[width=1.00\textwidth, angle=0]{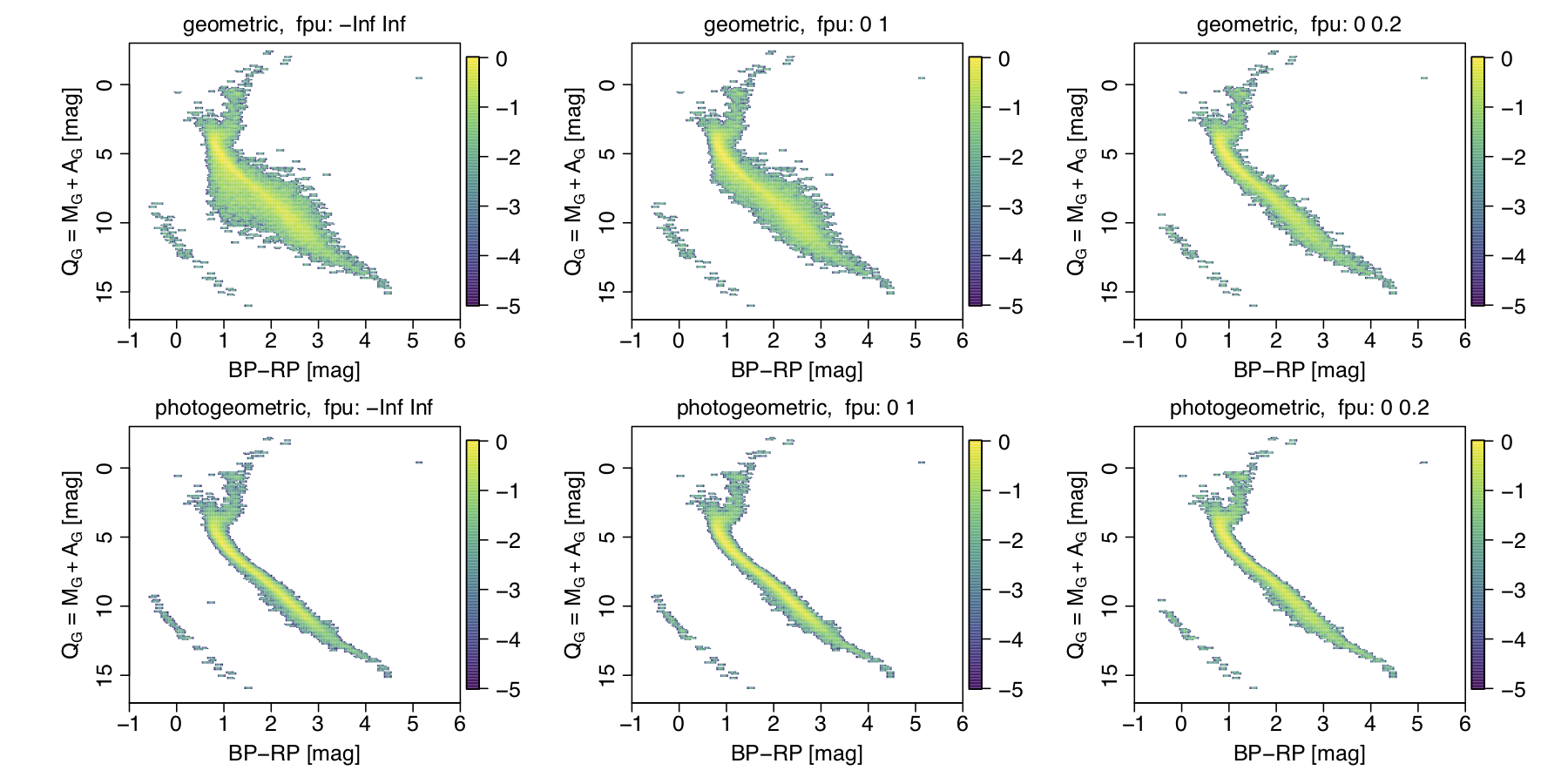}
\caption{The CQD inferred for mock catalogue HEALpixel 6200 using the median geometric distance (top row) and median photogeometric distance (bottom row) for three ranges of the true fractional parallax uncertainty (fpu): all (left), 0--1.0 (middle) and 0--0.2 (right). The colour scale is a logarithmic (base 10) density scale relative to the highest density cell in each panel.
\label{fig:mock_inference_CQDs_1}}
\end{center}
\end{figure*}

\begin{figure*}
\begin{center}
\includegraphics[width=1.00\textwidth, angle=0]{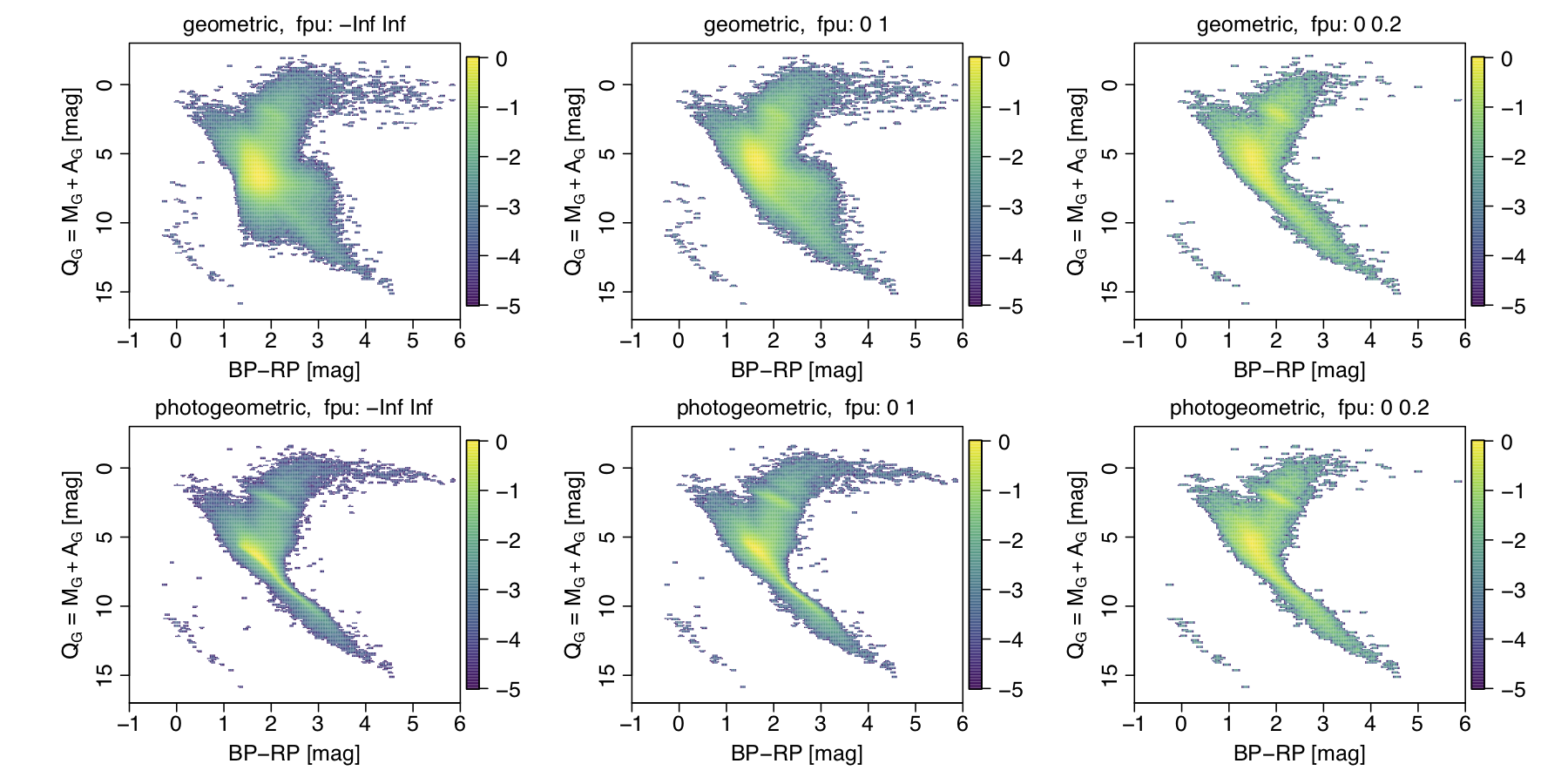}
\caption{As Figure~\ref{fig:mock_inference_CQDs_1} but now for HEALpixel 7593.
\label{fig:mock_inference_CQDs_2}}
\end{center}
\end{figure*}

We can also assess the quality of our distance estimates by computing $\QG \,=\, \gmag - 5\logten\rmed + 5$ and plotting the resulting CQD. We do this for both the geometric and photogeometric distances, for three ranges of fpu, for HEALpixel 6200 in Figure~\ref{fig:mock_inference_CQDs_1} and HEALpixel 7593 in Figure~\ref{fig:mock_inference_CQDs_2}. These can be compared to the CQD for the same HEALpixels constructed using the true distances shown in Figure~\ref{fig:mockCQD}. Imperfect distance estimates can only move sources vertically in this diagram as the \bprp\ colours are not changed. We see how the inferred main sequence is wider for the larger fpu samples for the geometric distances (left two columns in both plots), but much less so for the photogeometric distances. This is again due to the stablizing influence of the \QG\ prior. 
Both distance estimates are able to recover the primary structures: the main sequence, white dwarf sequence, giant branch, and horizontal branch.
  These plots will be useful when it comes to analysing the results on the real \release\ data, because they do not involve the truth as a reference.

\vspace*{1em}
\section{Analysis of distance results in \release}\label{sec:results}

We applied our inference code (written in R) to the 1.47 billion sources in \gaia\ \release\ that have parallaxes. This required $1.6\times 10^{12}$ evaluations of the posteriors and took 57\,000 CPU-core-hours.
Throughout this section the term ``fpu" of course refers to the {\em measured} fractional parallax uncertainty, as we do not know the true parallax.

\subsection{Analysis of two HEALpixels}\label{sec:analysis_two_healpixels}

\subsubsection{Distance distributions and uncertainties}

\begin{figure*}
\begin{center}
\includegraphics[width=1.0\textwidth, angle=0]{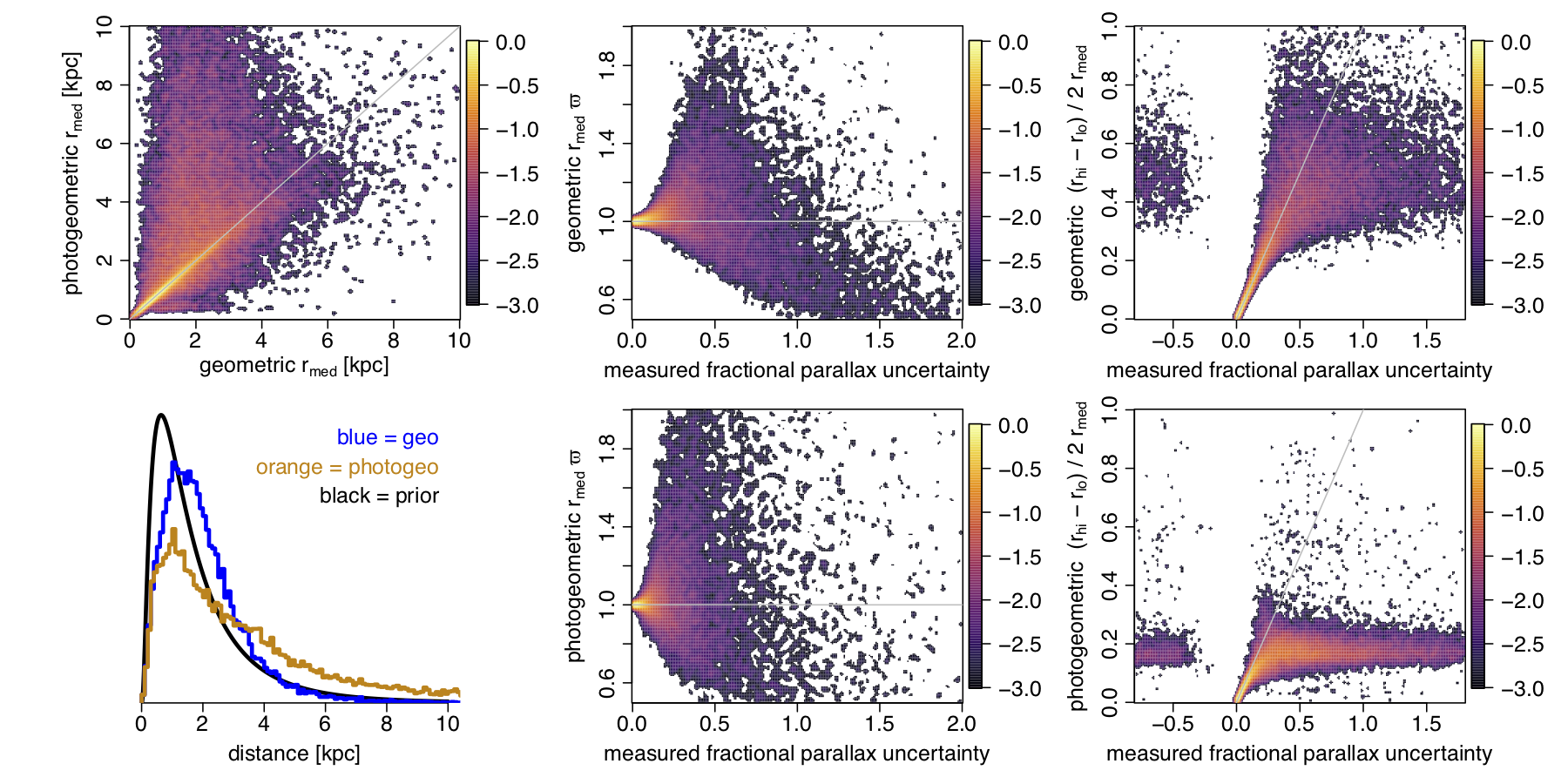}
\caption{\release\ distance results for HEALpixel number 6200 at 
$(\glon, \glat) = (285.7\degree, 34.8\degree)$. 
The colour scale in the density plots is logarithmic (base 10) relative to the highest density cell in each panel. 
The top-left panel compares the median geometric and photogeometric distances. 
The bottom-left panel shows normalized histograms on a linear scale of the median geometric (blue) and photogeometric (orange) distances, compared to the distance prior (black).
The middle column shows the ratio of the inferred distance to the inverse parallax distance 
as a function of the measured fractional parallax uncertainty (fpu). Note that the apparent lack of sources in the lower panel at fpus above about 1.0 is mostly a plotting artefact: regions with too-low a density of sources are white.
The two panels in the right column show the fractional symmetrized distance uncertainty also as a function of fpu (note the different scales).  This plot is available for all HEALpixels with the auxiliary information online.
\label{fig:real_inference_1}
}
\end{center}
\end{figure*}

\begin{figure*}
\begin{center}
\includegraphics[width=1.0\textwidth, angle=0]{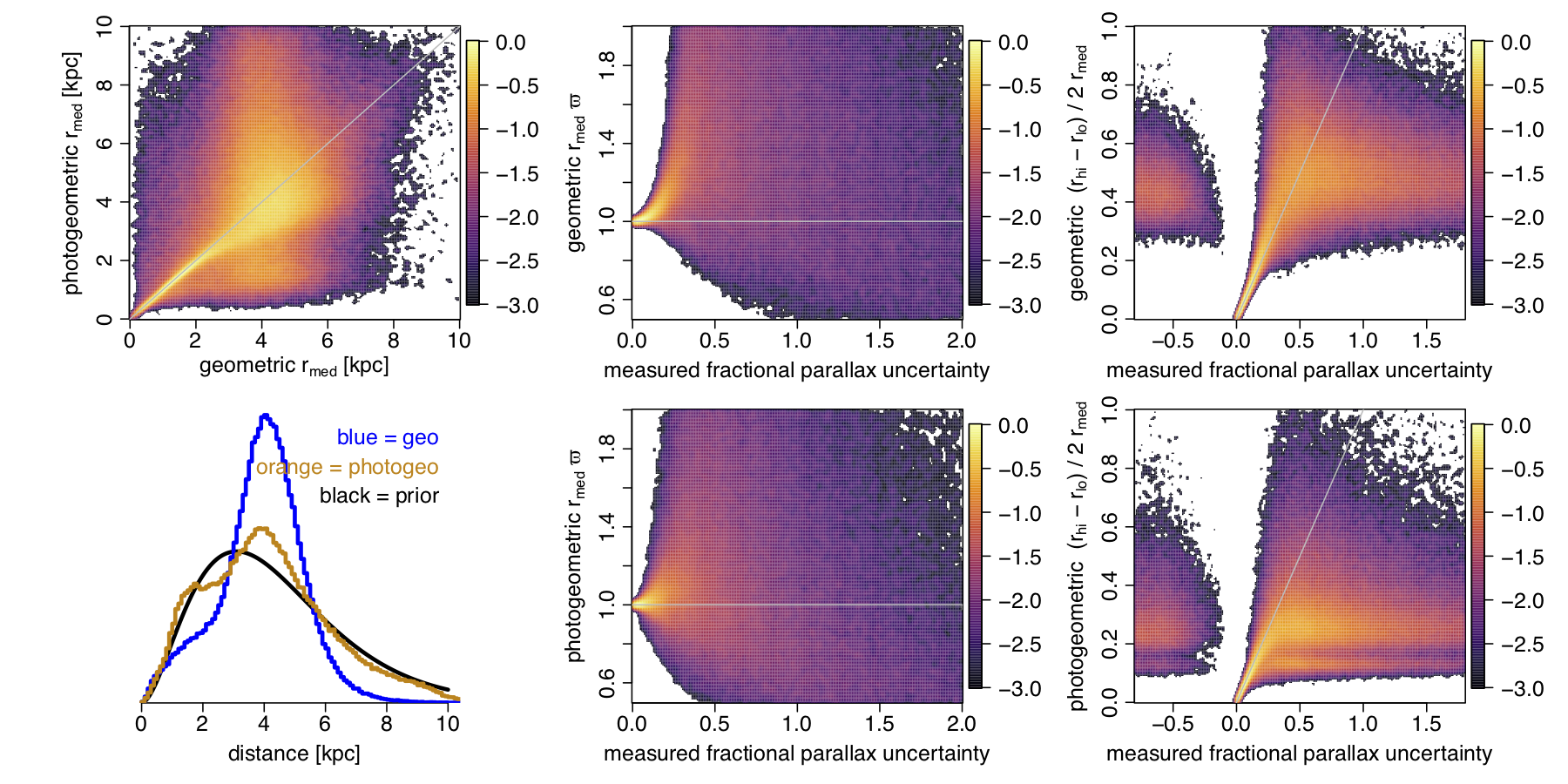}
\caption{As Figure~\ref{fig:real_inference_1} but for  HEALpixel number 7593 at $(\glon, \glat) = (29.0\degree, 7.7\degree)$. 
\label{fig:real_inference_2}
}
\end{center}
\end{figure*}

Results for our two example HEALpixels are shown in Figures~\ref{fig:real_inference_1} and~\ref{fig:real_inference_2}.
The two panels in the left column compare the two types of distance estimates. As expected, the photogeometric estimates extend to larger distances (see section~\ref{sec:mock_qualitative_analysis} for an explanation).
The middle columns plot the ratio of the inferred distance to the inverse parallax distance (corrected for the zeropoint). The latter is of course generally a poor measure of distance because it is not the true parallax, and this is the whole point of using an appropriate prior (see section~\ref{sec:introduction} and references therein). We see that both of our distance estimates converge to $1/\parallax$ in the limit of small fpu. 
Although the apparent lack of sources at large fpu in the lower middle panels is primarily a plotting artefact (due to the finite density scale), the two samples in the upper and lower panels are not identical, because not all sources have photogeometric distances.
For HEALpixel 6200 there are 24\,007 sources with geometric distances and  23\,829 with photogeometric distances. For HEALpixel 7592 these numbers are  385\,902 and 369\,608 respectively.

The panels in the right columns of Figures~\ref{fig:real_inference_1} and~\ref{fig:real_inference_2} show how the fractional symmetrized  distance uncertainty varies with fpu. At small (positive) fpu they are nearly equal for both geometric and photogeometric distances, because here the likelihood dominates the posterior. At larger fpu the geometric distances become more uncertain, which is commensurate with their lower expected accuracy. 
For very large fpu ($\gg 1$) the geometric distances and their uncertainties will be dominated by the prior, which for HEALpixel 7593 has a median of 3.98\,kpc and lower (16th) and upper (84th) quantiles of 2.06\,kpc and 6.74\,kpc respectively (corresponding to a fractional distance uncertainty of 0.59).
The photogeometric fractional distance uncertainties tend to be smaller than the geometric ones. This is because the \QG\ prior (section~\ref{sec:QG_prior}) is usually more informative than the distance prior. 

We extend the axes in the right panels of Figures~\ref{fig:real_inference_1} and~\ref{fig:real_inference_2} to negative fpu, which occur when sources have negative parallaxes. One of the advantages of probabilistic inference is to provide meaningful distances for negative parallaxes (a quarter of all parallaxes in EDR3).
Negative observed parallaxes ususally correspond to sources with small true parallaxes,
and although such measurements generally have reduced impact on the posterior, they do carry information.
They do not yield precise distances, but insofar as the prior can be trusted the posterior and resulting confidence intervals are meaningful. We see from the figures that the precisions are low for both types of distance, but sometimes more constrained for the photogeometric ones due to the additional use of colour and magnitude.  In some senses the negative fpu regime is a continuation of the $\fpu >> 1$ regime (see Figures 3 and 6 of paper II).

\subsubsection{Colour--\QG\ diagrams}

\begin{figure*}[p]
\begin{center}
\includegraphics[width=1.00\textwidth, angle=0]{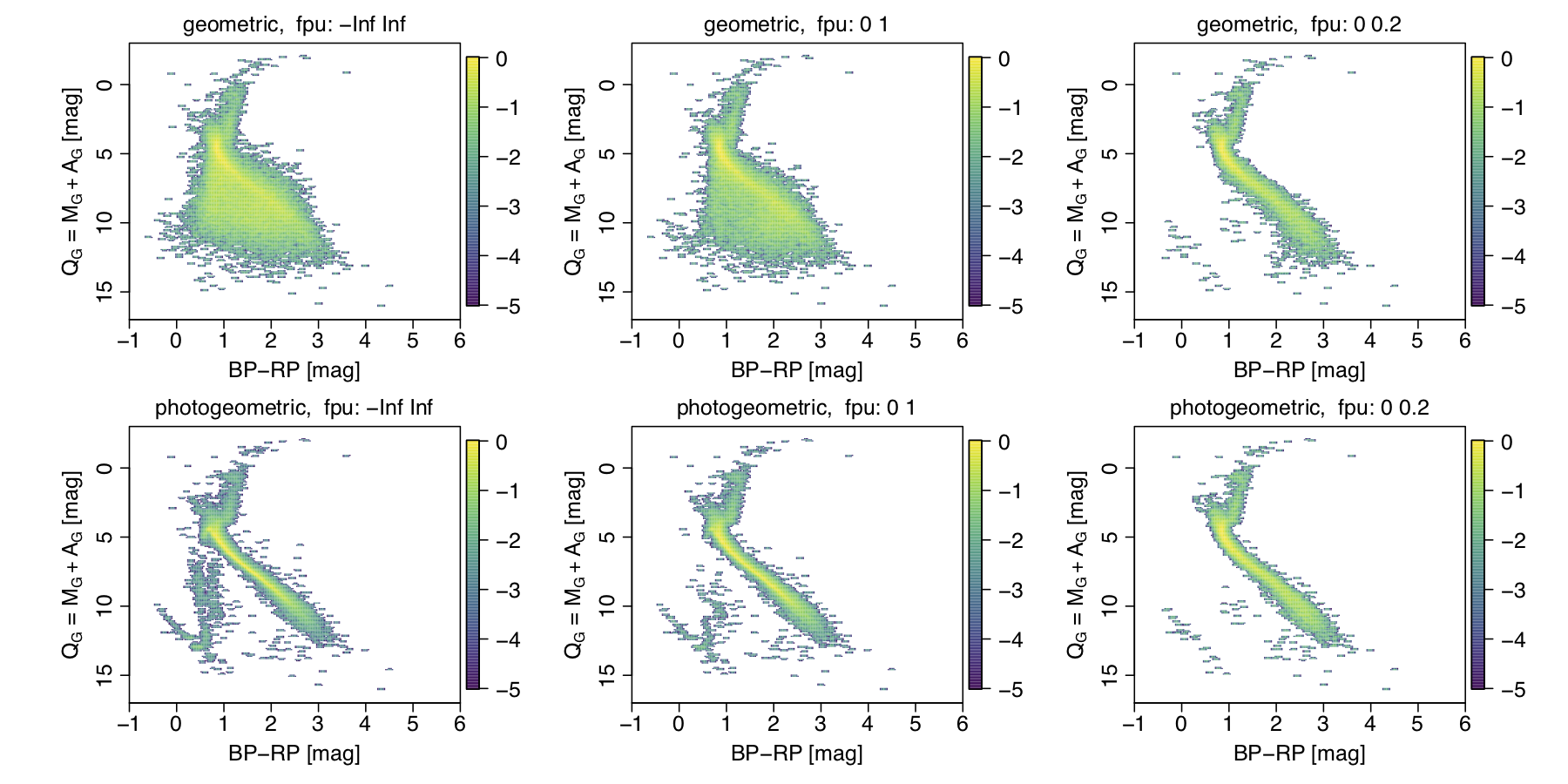}
\caption{The CQD inferred for \release\ HEALpixel 6200 using the median geometric distance (top row) and median photogeometric distance (bottom row) for three ranges of the measured fpu: all (left), 0--1.0 (middle) and 0--0.2 (right).
In total there are 24\,007 sources with geometric distances and 23\,829 with photogeometric distances.
No other filtering has been applied. The colour scale is a logarithmic (base 10) density scale relative to the highest density cell in each panel, so is not comparable across panels.
This plot (including also a comparison with the prior CQD) is available for all HEALpixels with the auxiliary information online.
\label{fig:real_inference_CQDs_1}}
\end{center}
\end{figure*}

\begin{figure*}
\begin{center}
\includegraphics[width=1.00\textwidth, angle=0]{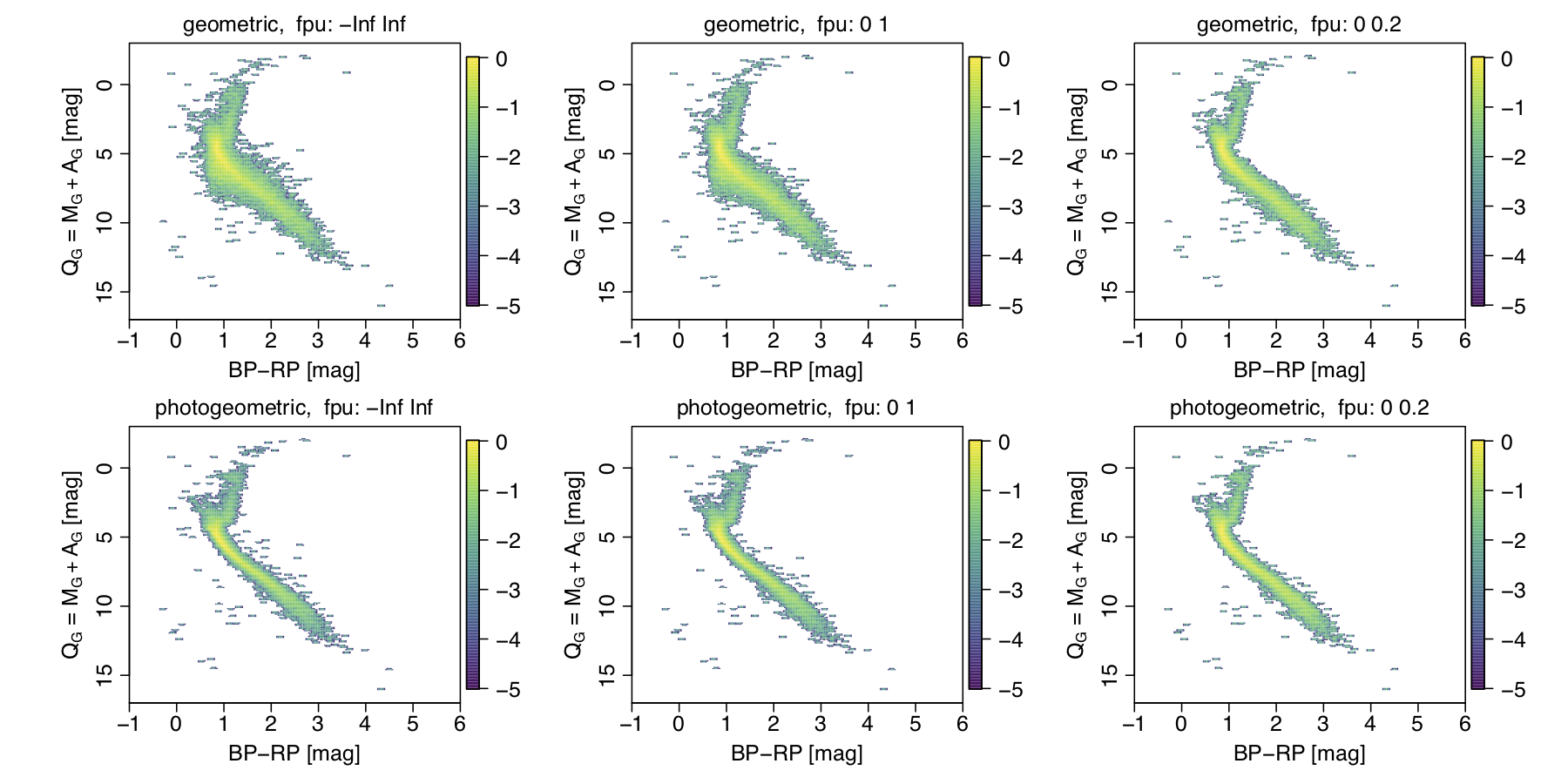}
\caption{As Figure~\ref{fig:real_inference_CQDs_1} but now excluding the 54\% of sources in this HEALpixel with $\gmag > 19.0$\,mag.
\label{fig:real_inference_CQDs_1b}}
\end{center}
\end{figure*}

From the inferred median distances we can compute the median \QG\ via equation~\ref{eqn:continuity} and then plot the CQD. This is shown in Figure~\ref{fig:real_inference_CQDs_1} for HEALpixel 6200 for the geometric distance (top row) and photogeometric distance (bottom row) for three different ranges of the fpu.  
As interstellar extinction should be low towards this high latitude field (around 0.15\,mag in \mock),
$\QG \simeq \MG$ so this CQD is similar to the colour-absolute magnitude diagram.
In all of the panels we see a well-defined main sequence and giant branch, as well as a white dwarf sequence in some of the panels.
Comparing the upper and lower panels we  see how the photogeometric distances constrain the \QG\ distribution more than the geometric distance do. 
The puffing-up of the geometric CQD is due to sources with large fpu: their distances tend to be underestimated (see section~\ref{sec:mock_qualitative_analysis}) so \QG\ becomes larger -- intrinsically fainter -- for a given \gmag\ (see equation~\ref{eqn:continuity}).
This puffing-up diminishes as we successively reduce the range of fpu, as shown in the middle and right columns of Figure~\ref{fig:real_inference_CQDs_1}.

The photogeometric CQD for the full fpu range (bottom left panel of Figure~\ref{fig:real_inference_CQDs_1}) shows a conspicuous blob of sources at $\bprp \simeq 0.5$\,mag between the MS and WD sequences. 
These are sources with spuriously large parallaxes, well known from \gdr{2} \citep{2018A&A...616A..17A} and still present, if less so, in \release\ \citep{fabricius_etal_gedr3_validation, gedr3_nearbystars}.
They are usually close pairs of sources that receive incorrect astrometric solutions, as the \release\ astrometric model is only suitable for single stars \citep{lindegren_etal_gedr3_astrometry}. Figure~\ref{fig:real_inference_CQDs_1} shows
that spurious parallaxes are less common among the smaller fpu subsample. 
The \QG\ prior will often help to constrain the distance of these spurious solutions and thus place them on the correct part of the CQD. This is only partially successful at around $\bprp \simeq 0.5$\,mag in this HEALpixel, however, because the distance prior may still be pulling truly very distant sources with larger fpu towards us.

Sources with spurious parallaxes are preferentially faint.
To quote from \cite{gedr3_release}:
``For faint sources ($\gmag > 17$ for 6-p astrometric solutions and $\gmag > 19$ for 5-p solutions) and in crowded regions the fractions of spurious solutions can reach 10 percent or more."
This can be seen in Figure~\ref{fig:real_inference_CQDs_1b}
where we replot the CQD only for sources with $\gmag < 19.0$\,mag. This also reduces the puffing-up of the geometric CQD, although some of this reduction is simply because magnitude is correlated with fpu, so a magnitude cut also lowers the fpu. 

\begin{figure*}
\begin{center}
\includegraphics[width=1.00\textwidth, angle=0]{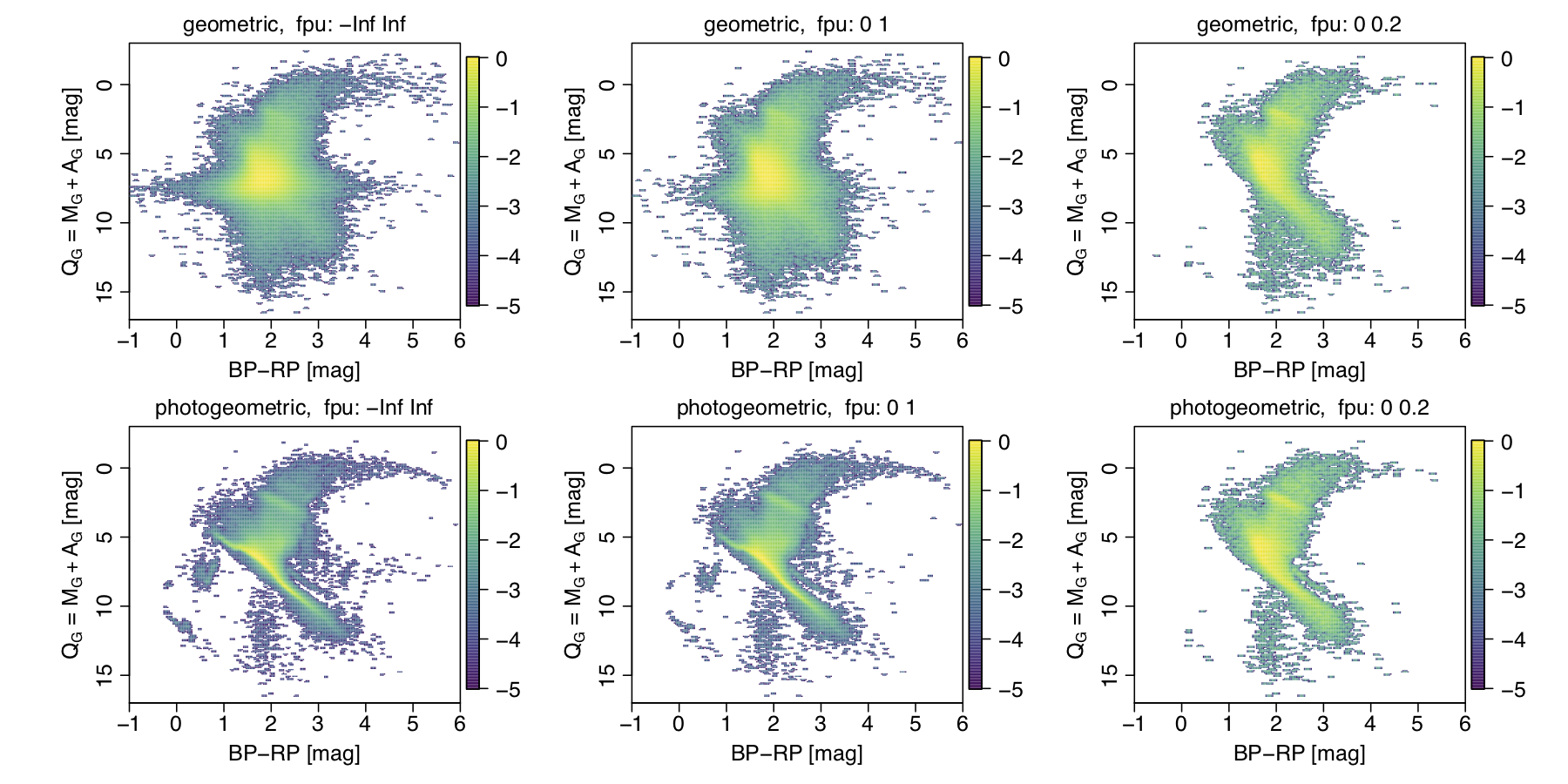}
\caption{As Figure~\ref{fig:real_inference_CQDs_1} but now for HEALpixel 7593. All sources are shown (no magnitude cut).
In total there are 385\,902 sources with geometric distances and 369\,608 with photogeometric distances.
\label{fig:real_inference_CQDs_2}}
\end{center}
\end{figure*}

\begin{figure*}
\begin{center}
\includegraphics[width=1.00\textwidth, angle=0]{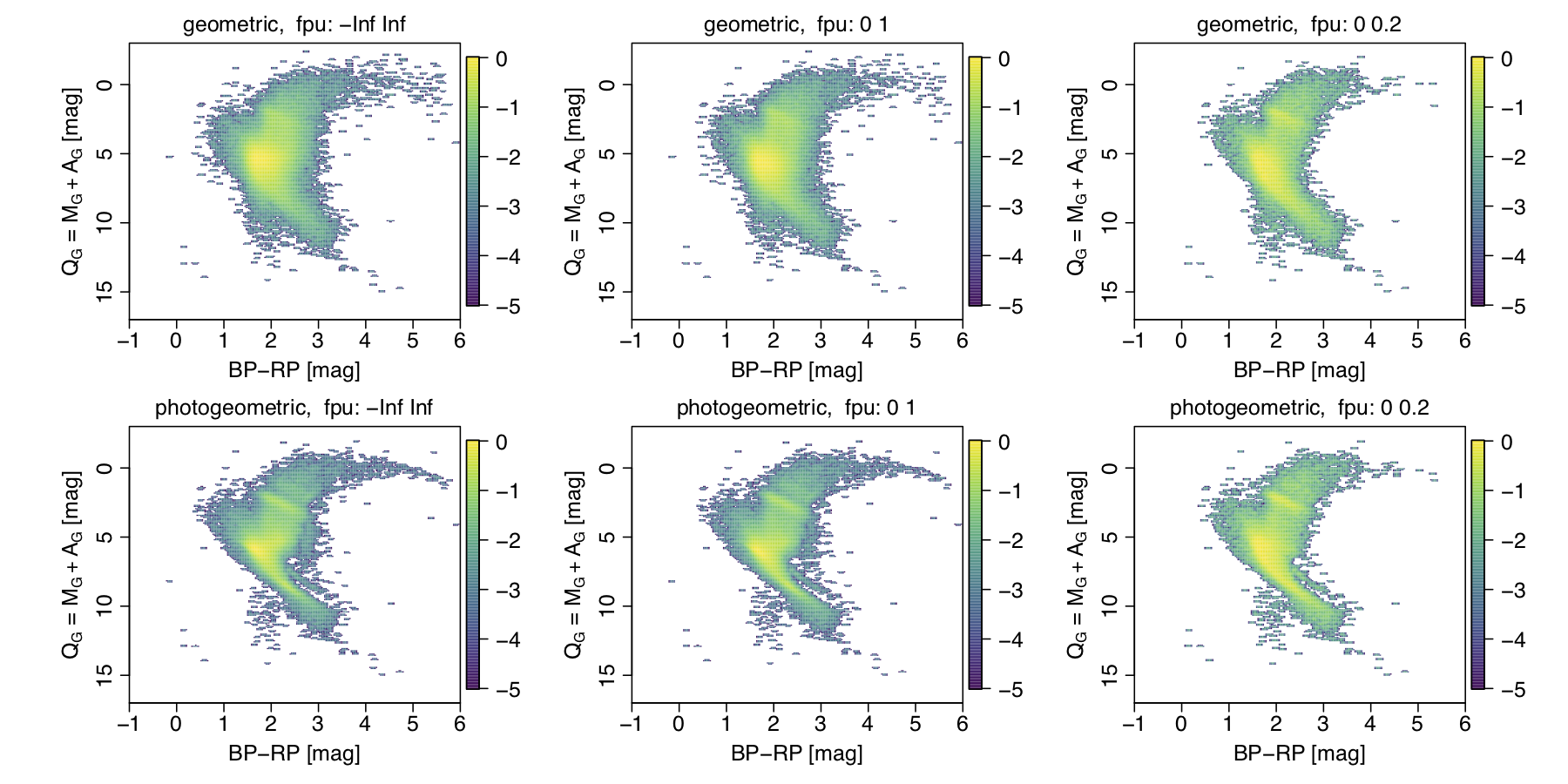}
\caption{As Figure~\ref{fig:real_inference_CQDs_2} but now excluding the 70\% of sources in this HEALpixel with $\gmag > 19.0$\,mag to remove spurious sources.
\label{fig:real_inference_CQDs_2b}}
\end{center}
\end{figure*}

These effects can be seen more prominently in the low latitude HEALpixel 7593, shown in Figures~\ref{fig:real_inference_CQDs_2} and~\ref{fig:real_inference_CQDs_2b}. 
Due to the larger mean distance of stars at low latitudes (see section~\ref{sec:mock_quantitative_analysis}), as well as the more complex stellar populations and larger mean extinction (up to 3.5\,mag),
the CQD is more complex. For the full fpu range, the geometric CQD in Figure~\ref{fig:real_inference_CQDs_2} is quite washed out, due in part to large fpus and spurious parallaxes, although an extincted red clump is visible. The photogeometric CQDs are cleaner, with a better defined main sequence. 
The CQD for the $\gmag < 19.0$\,mag subsample (Figure~\ref{fig:real_inference_CQDs_2b}) again shows the removal of spurious sources.
Section 3.2 
of \cite{fabricius_etal_gedr3_validation} analyses spurious astrometric solutions and offers more sophisticated ways of identifying them than a simple magnitude cut.

\subsection{All sources}

We now look at a representative sample of the entire catalogue.
All plots and analyses in this section use a random selection of 0.5\% of all sources from each HEALpixel. This has 7\,344\,896 geometric and 6\,739\,764 photogeometric distances.

\begin{figure}
\begin{center}
\includegraphics[width=0.50\textwidth, angle=0]{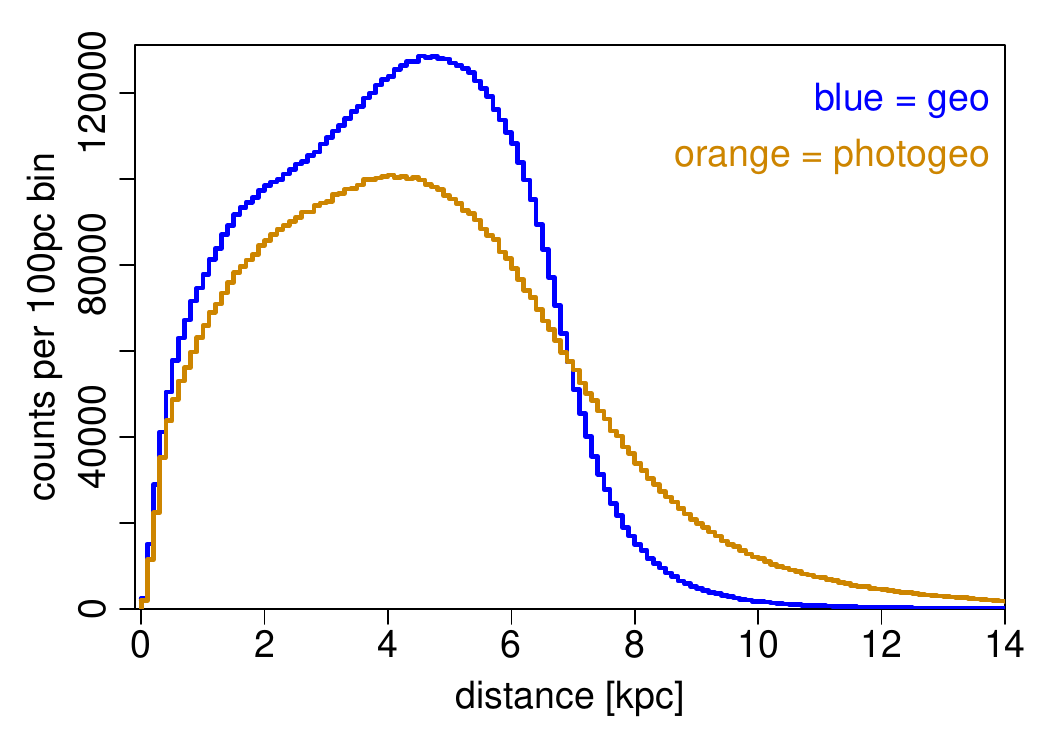}
\caption{Distribution of inferred geometric and photogeometric median distances, \rmed, in \release. This plot uses a random sample of 0.5\% of all sources in each HEALpixel.
\label{fig:distance_histogram_allhp}}
\end{center}
\end{figure}


\begin{figure*}
\begin{center}
\includegraphics[width=1.00\textwidth, angle=0]{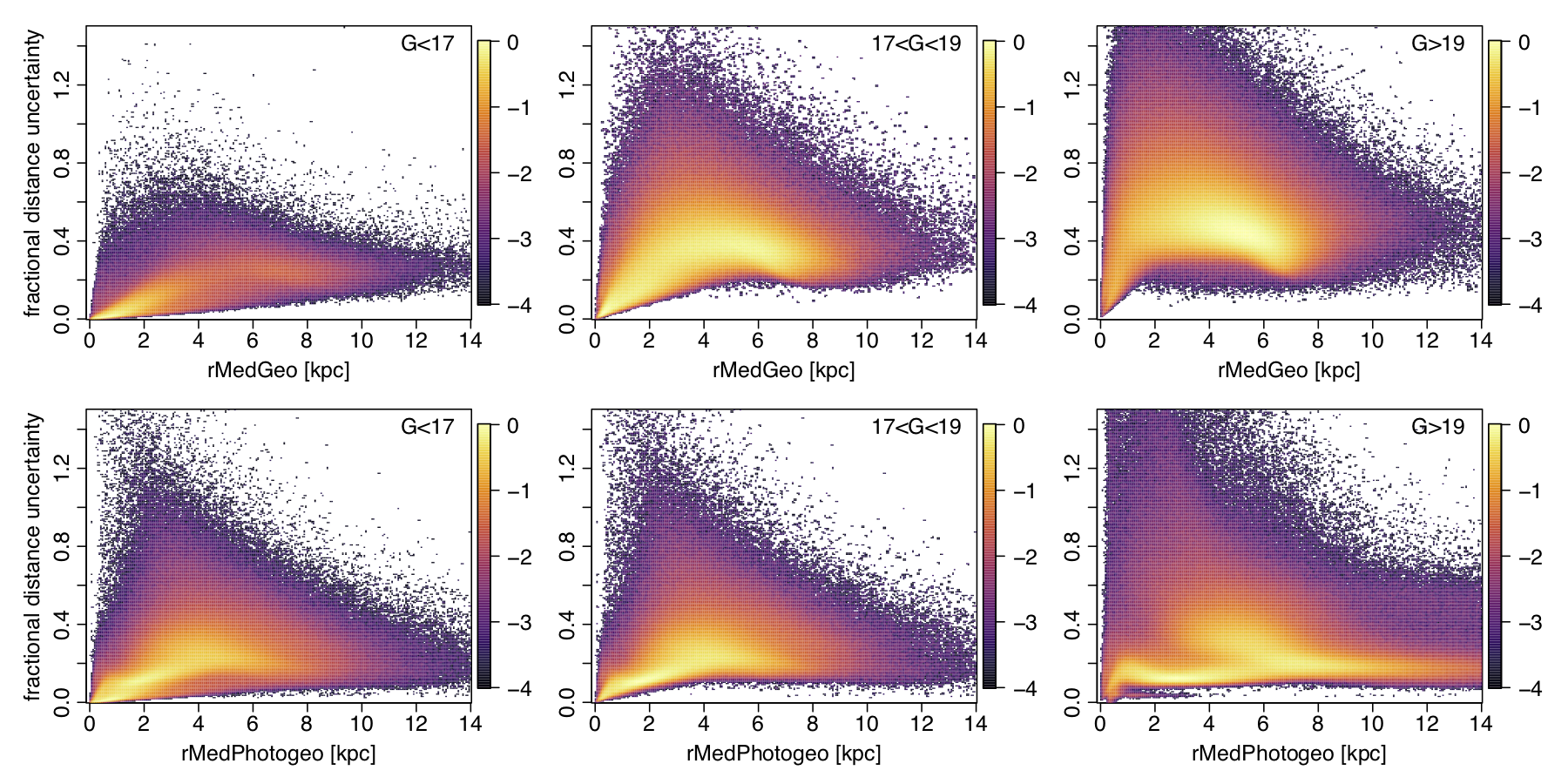}
\caption{Fractional symmetrized distance uncertainty, $(\rhi-\rlo)/2\rmed$, vs distance for the geometric distance estimates (top) and photogeometric distance estimates (bottom) for the three different \gmag\ ranges.
The colour scale is a logarithmic density (base 10) scale relative to the highest density cell in each panel.
This plot uses a random sample of 0.5\% of all sources in each HEALpixel. 
\label{fig:fdu_vs_distance_allhp}}
\end{center}
\end{figure*}

Figure~\ref{fig:distance_histogram_allhp} shows the distribution of distances. As expected, the photogeometric distances extend to larger distances that the geometric one.
The fractional symmetrized distance uncertainties as a function of distance are shown in Figure~\ref{fig:fdu_vs_distance_allhp} for three different magnitude ranges. 
As noted earlier, the photogeometric distance uncertainties are generally smaller than the geometric ones, at least for fainter sources. This plot also shows again that photogeometric estimates extend to larger distances.

\subsubsection{Colour--\QG\ diagrams}

\begin{figure}
\begin{center}
\includegraphics[width=0.50\textwidth, angle=0]{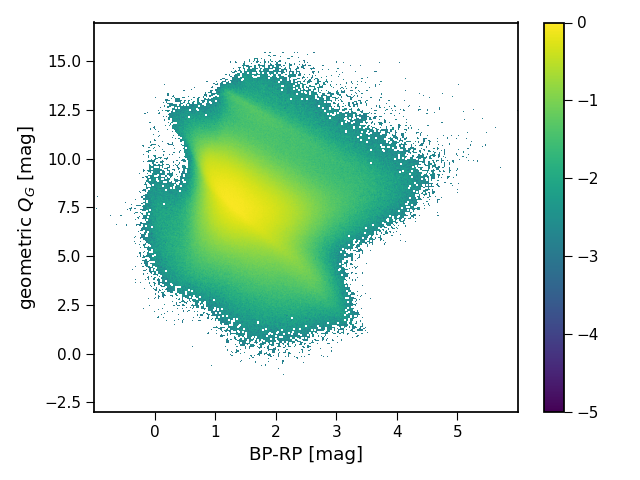}
\includegraphics[width=0.50\textwidth, angle=0]{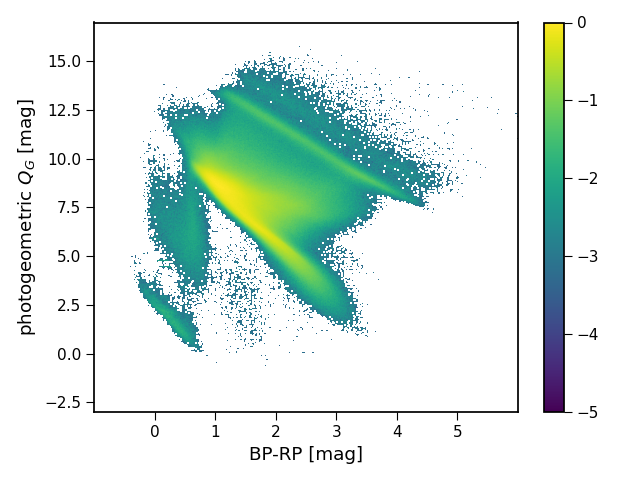}
\caption{The \release\ CQD over the whole sky using the geometric distances (top) and photogeometric distances (bottom). This plot uses a random sample of 0.5\% of all sources in each HEALpixel. These plots include sources of all magnitude and fpu, and so include sources with spurious parallaxes.
\label{fig:CQG_allsky}
}
\end{center}
\end{figure}

Figure~\ref{fig:CQG_allsky} shows the CQD over the whole sky. Because the sample is a constant random fraction per HEALpixel it is numerically dominated by sources at low latitude Galactic latitudes where there can be significant interstellar extinction. This is apparent from the upper diagonal feature --   especially clear in the photogeometric panel -- which is the red clump stretched by extinction/reddening. The white dwarf sequence appears clearly in the photogeometric CQD. Although some white dwarfs are correctly placed in the CQD by the geometric distances, they are not visible here due to the finite dynamic range of the plotted density scale. Furthermore, for reasons explained in section~\ref{sec:mock_qualitative_analysis}, faint nearby sources with large fpu tend of have their geometric distances overestimated and therefore their \QG\ underestimated, thereby pushing them up from the true white dwarf sequence. These plots have not filtered out spurious sources, some of which are clearly visible in the photogeometric CQD as the blob between the upper MS and the white dwarf sequence.
Other broad differences between the geometric and photogeometric CQDs were explained in section~\ref{sec:mock_inferered_CQDs}.

\subsubsection{Distribution on the sky}

\begin{figure}
\begin{center}
\includegraphics[width=0.50\textwidth, angle=0]{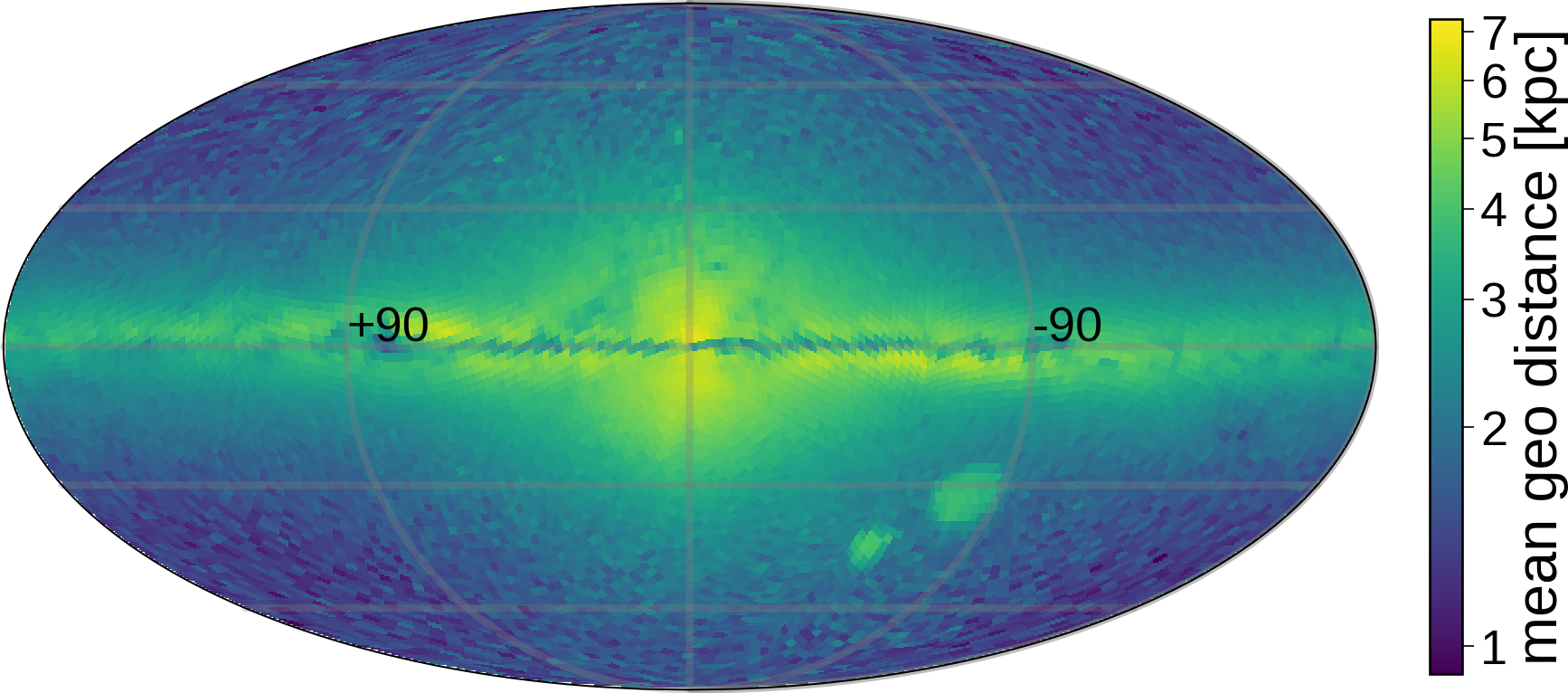}
\includegraphics[width=0.50\textwidth, angle=0]{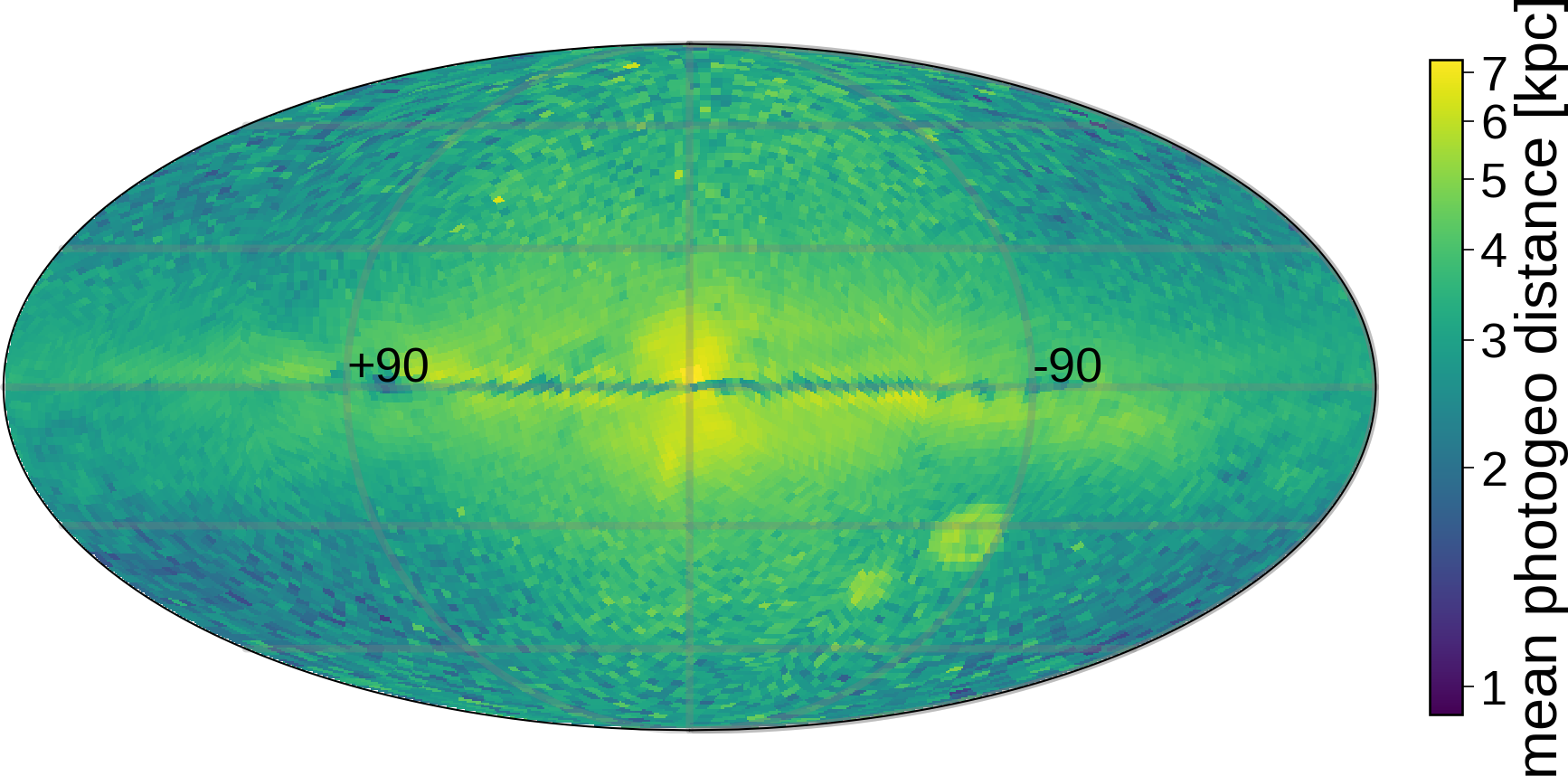}
\includegraphics[width=0.50\textwidth, angle=0]{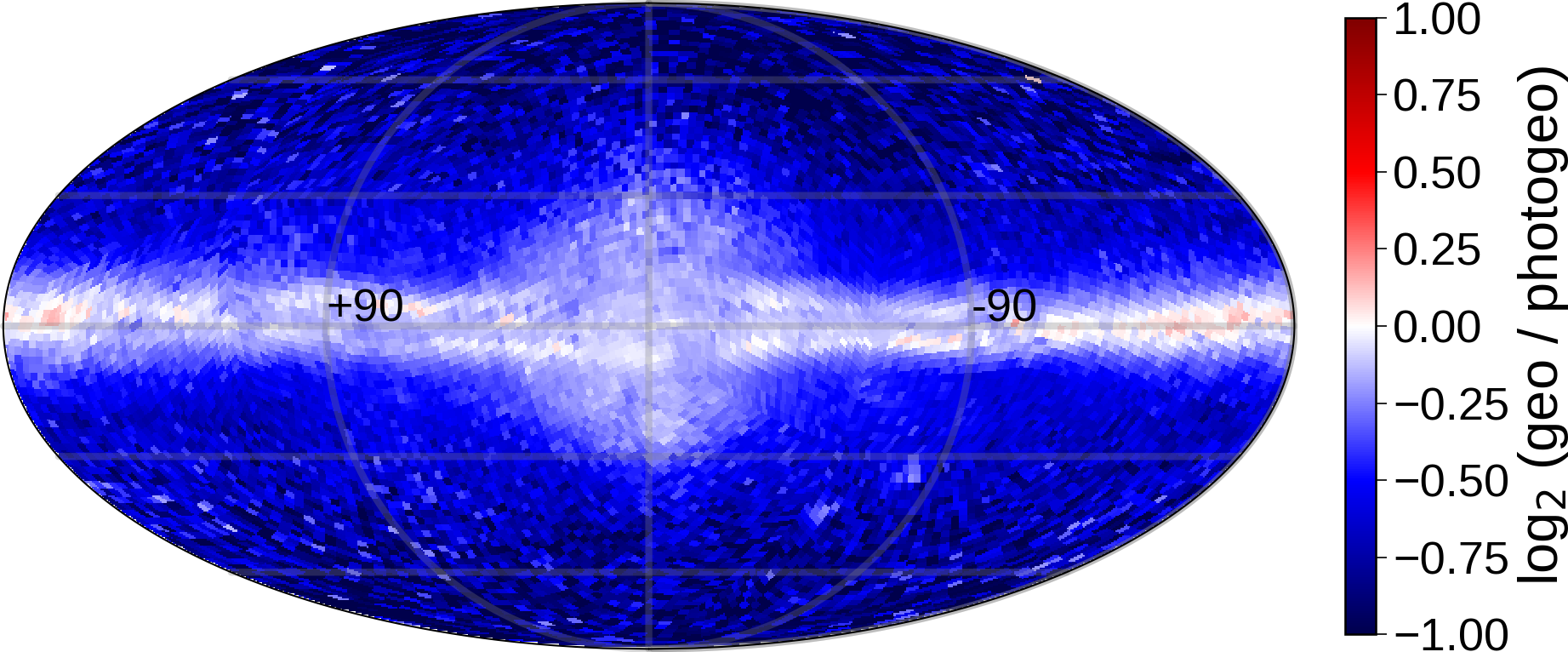}
\caption{The mean distance of sources per HEALpixel (level 5) for our median geometric distances (top) and median photogeometric distances (middle), and the log$_2$ ratio of these (bottom), i.e. log$_2$(geo/photogeo).
This plot uses a random sample of 0.5\% of all sources in each HEALpixel.
\label{fig:distance_allsky}}
\end{center}
\end{figure}

Figure~\ref{fig:distance_allsky} shows the mean distance of sources (i.e.\ mean of \rmed) in each HEALpixel in our catalogue, as well as the ratio of these in log base 2. 
Over all HEALpixels the 5th, 50th, and 95th percentiles of the mean of the geometric distances are  1.3, 2.1, and 4.4\,kpc respectively. The percentiles for the mean of the photogeometric distances are 2.2, 3.3, and 5.0\,kpc. These translate into low ratios of geometric to photogeometric distances in general. Only in the Galactic plane and the bulge are the two mean distances comparable. At high Galactic latitudes the photogeometric average is easily twice as large as the geometric average. 

\subsubsection{Galactic spatial distribution}

\begin{figure*}
\begin{center}
\includegraphics[height=0.40\textheight, angle=0]{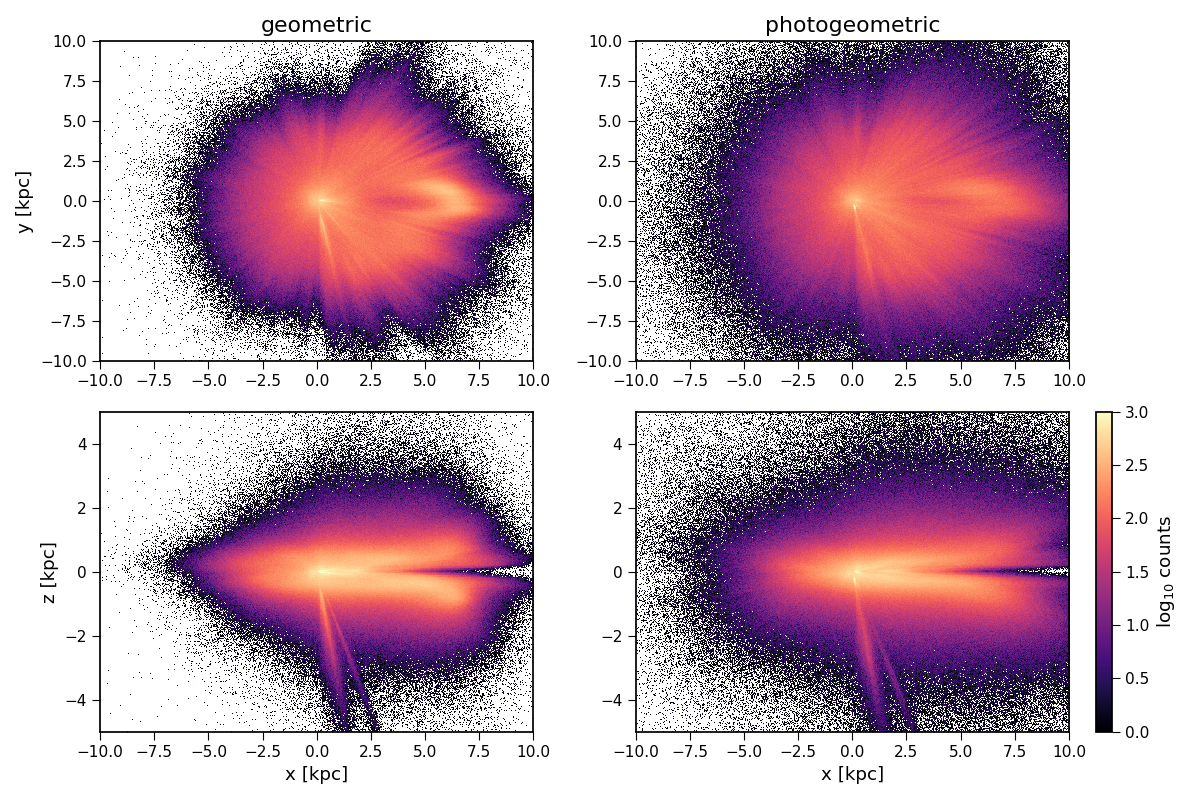}
\caption{Projected distribution of \release\ stars in the Galaxy using our geometric distances (left) and photogeometric distances (right). The projections are in Galactic Cartesian coordinates with the Sun at the origin. 
The Galactic North Pole is in the positive $z$ direction and the Galactic centre is at around $(+8,0,0)$\,kpc. Galactic longitude increase anticlockwise from the positive x-axis. 
The top plots are the view from the Galactic North Pole. 
The bottom plots are a side view. 
This plot uses a random sample of 0.5\% of all sources in each HEALpixel. 
\label{fig:distance_galprojection}
}
\end{center}
\end{figure*}

\begin{figure*}[p]
\begin{center}
\includegraphics[height=0.40\textheight, angle=0]{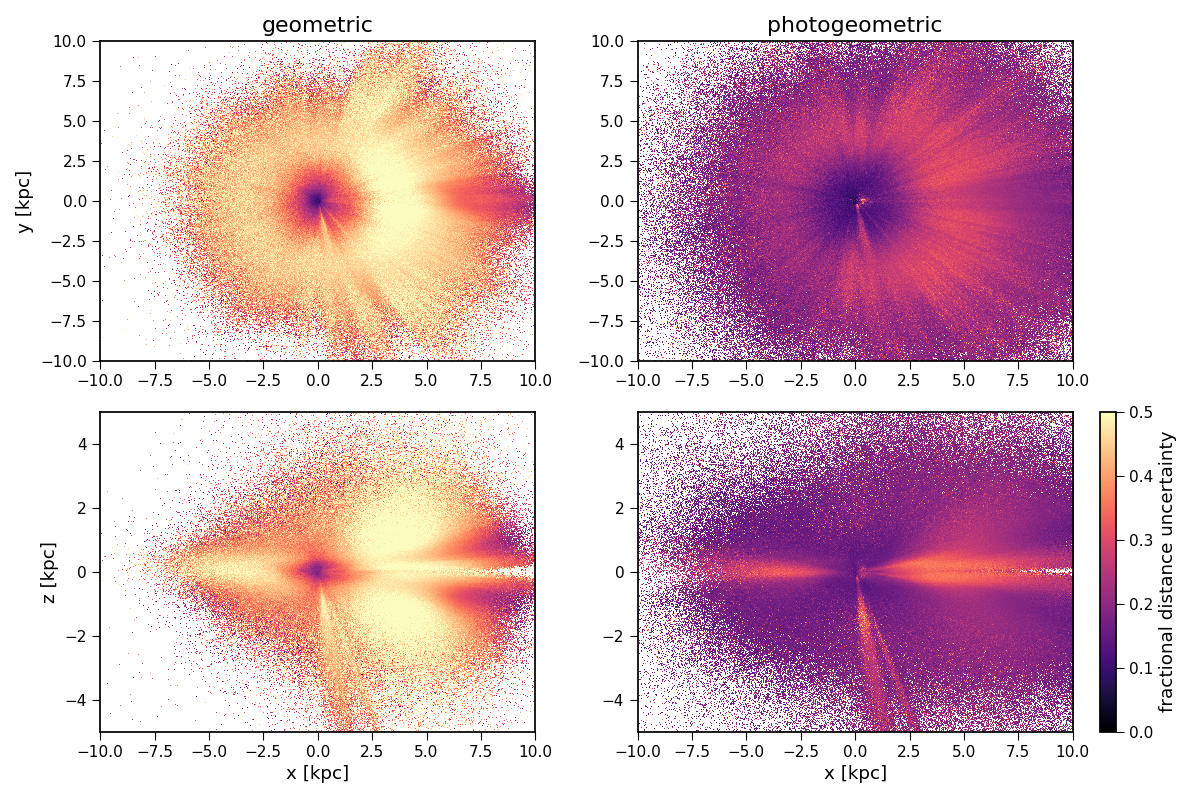}
\caption{As Figure~\ref{fig:distance_galprojection} but now showing the fractional symmetrized distance uncertainties, i.e.\  $(\rhi-\rlo)/2\rmed$.
\label{fig:distance_uncertainty_galprojection}}
\end{center}
\end{figure*}

Figure~\ref{fig:distance_galprojection} shows the projected distribution of stars in \release\ in the Galaxy using our distance estimates. The Sun is at the origin, and we see the expected larger density of sources in the first and fourth Galactic quadrants.
Finer asymmetries in the distribution projected onto the Galactic plane (upper panels) are presumably due to both a genuine asymmetry in the Galactic population and \gaia's scanning law. 
These, as well as nearby dust clouds, also explain the various radial lines pointing out from the origin.
The lack of sources in the fan around the positive x-axis in the lower panels is due to extinction in the Galactic plane. The overdensity in the same direction in the upper panels is the projection of the bulge.
The lower panels demonstrate the point made earlier (section~\ref{sec:mock_quantitative_analysis}) about being able to see sources to larger mean distances at lower Galactic latitudes. 

The high density rays extending below the Galactic plane (lower panels of Figure~\ref{fig:distance_galprojection}) are in the directions of the Magellanic Clouds. Many stars in these satellite galaxies are in \release\ -- they are some of the densest HEALpixels -- yet they are so far away (50--60\,kpc) that most have poor (and often negative) parallaxes, such that the inferred geometric distances are dominated by the prior (see appendix~\ref{sec:poor_parallax_limit} for further discussion). Our photogeometric distances are similarly poor, because we excluded the Magellanic Clouds from the mock CQD out of which our \QG\ priors are built. This was intentional: anyone interested in estimating distances to sources in the Magellanic clouds can do better than just use \gaia\ parallaxes and photometry.

Figure~\ref{fig:distance_uncertainty_galprojection} shows the fractional distance uncertainties also in Galactic projection. As expected, the uncertainties generally increase with distance from the Sun, but there are exceptions due to bright distant stars having more precise distances than faint nearby ones. The rays towards the Magellanic clouds also stand out as having larger uncertainties on the whole. 

\subsection{Validation using clusters}

\begin{figure*}
\begin{center}
\includegraphics[width=\textwidth, angle=0]{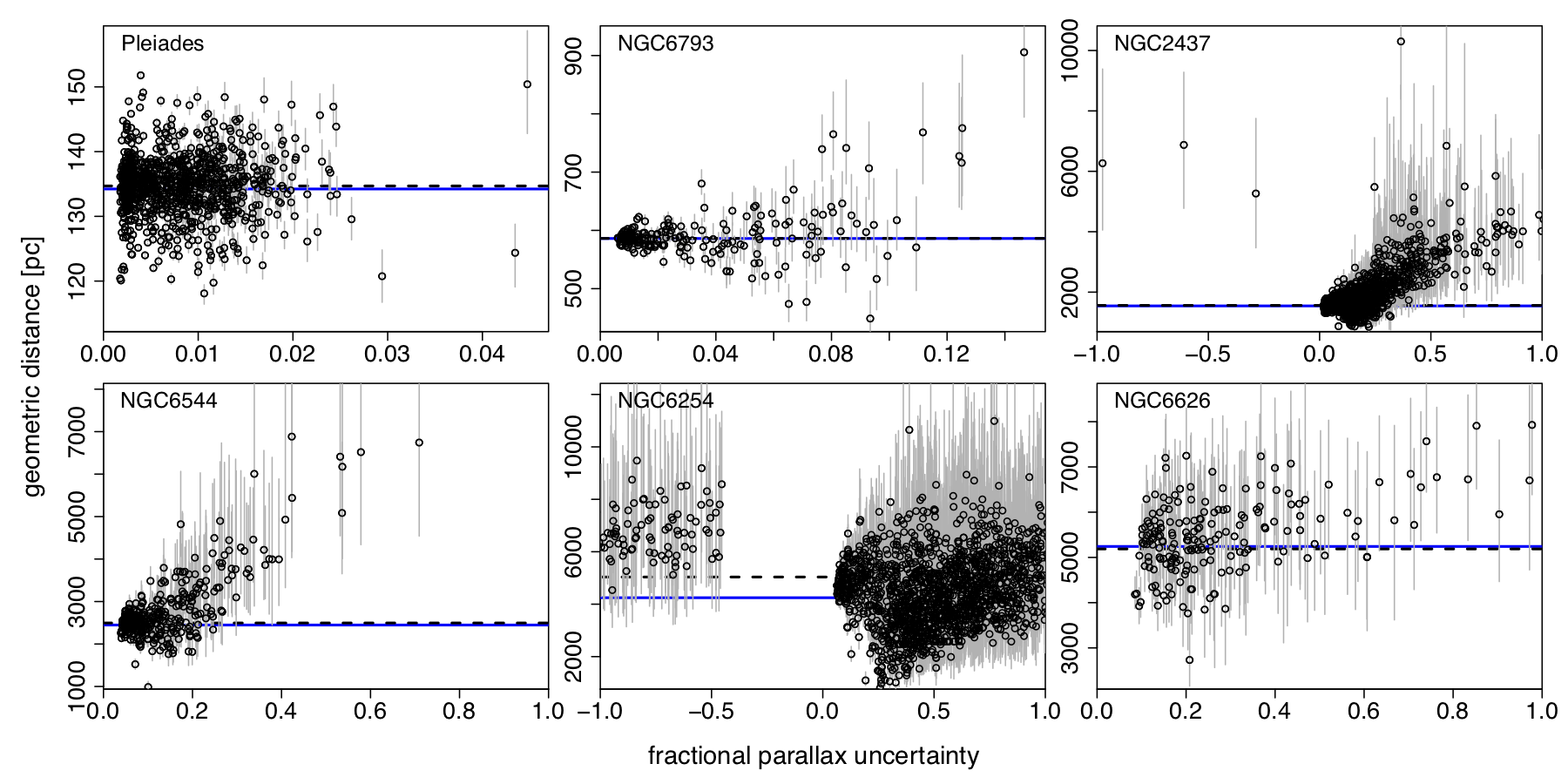}
\caption{Validation of the geometric distance estimates using star clusters (one per panel). Each panel shows the 
 estimated distance, \rmed, of the cluster members as open circles, as a function of the fractional parallax uncertainty \fpu. The error bars show the lower (\rlo) and upper (\rhi) bounds of the confidence intervals.
The distance range spans everything in the plotted fpu range,
but a few stars lie outside of the plotted fpu range for some clusters.
The dashed horizontal line is the inverse of the variance-weighted mean parallax for all cluster members (including any beyond the fpu limits plotted). The solid horizontal (blue) line is the weighted mean geometric distance for the same stars, where the weight is the inverse square of the symmetrized distance uncertainty. The clusters are ordered by increasing parallax distance.
\label{fig:cluster_validation_geo}
}
\end{center}
\end{figure*}

\begin{figure*}
\begin{center}
\includegraphics[width=\textwidth, angle=0]{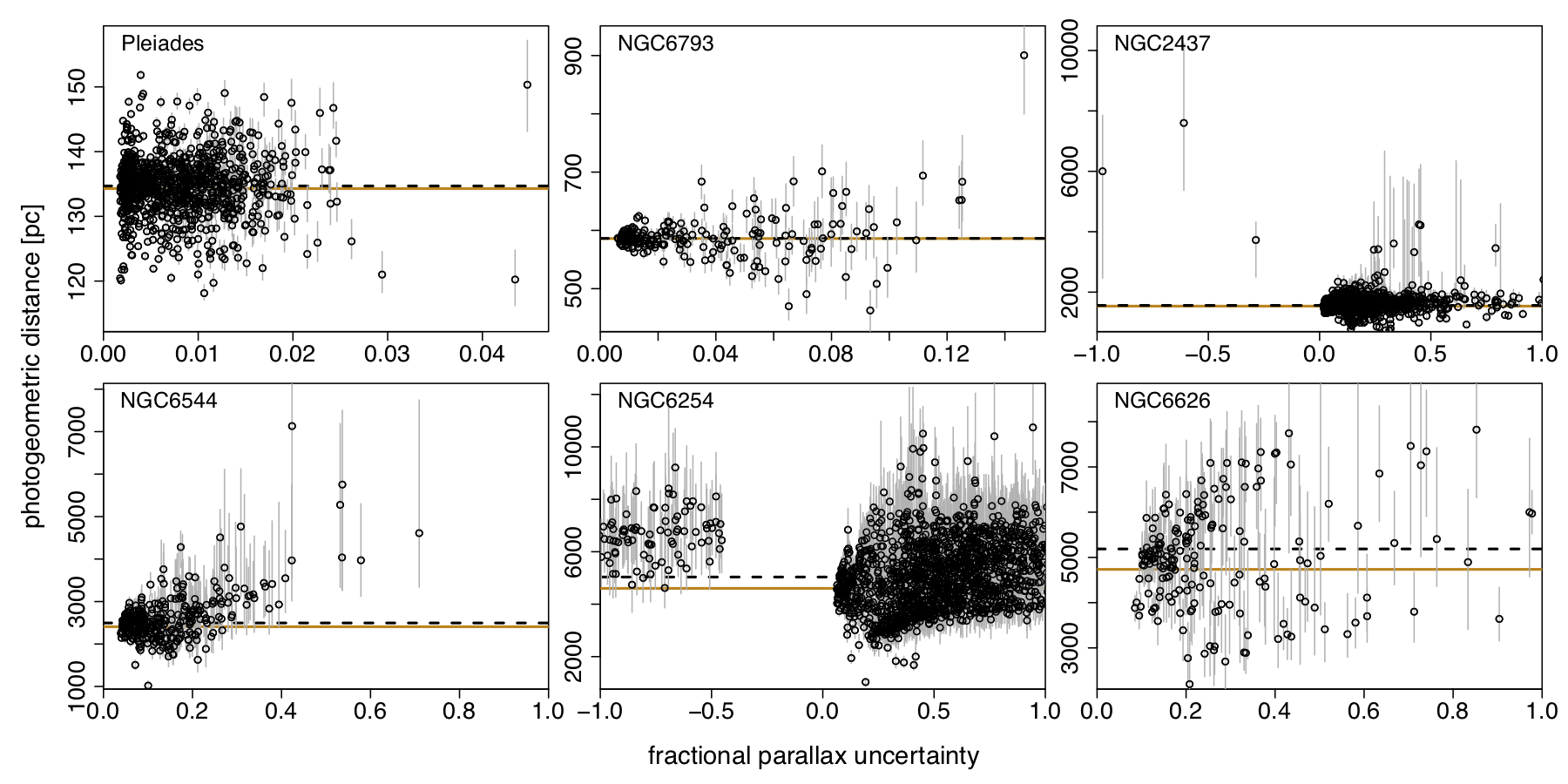}
\caption{As Figure~\ref{fig:cluster_validation_geo} but now for photogeometric distances. The solid horizontal (orange) line is the weighted mean photogeometric distance.
\label{fig:cluster_validation_photogeo}
}
\end{center}
\end{figure*}

Figures~\ref{fig:cluster_validation_geo} and~\ref{fig:cluster_validation_photogeo} show our geometric and photogeometric distances and their uncertainties for members of various star clusters. The membership lists have been drawn from paper IV. \object{NCG6254} (=\,\object{M10}) and \object{NGC6626} (=\,\object{M28}) are globular clusters; the rest are open clusters.
Recall that our prior does not include star clusters.
The horizontal dashed line in each panel
shows the inverse of the variance-weighted mean parallax of the members, i.e.\ a pure parallax distance for the cluster. Both of our distance estimates congregate around this for small, positive fpu, but deviate for large or negative fpu, as one would expect.
We generally see a larger deviation and/or scatter for the geometric distance: compare in particular the panels for \object{NGC2437} (=\object{M46}) and \object{NGC6254}.
Despite this, the weighted mean of our distances is often quite close to the pure parallax distance, even for clusters up to several kpc away.

We nevertheless emphasise that the inverse of the variance-weighted mean parallax will usually be a better estimate for the distance to a cluster than the mean of our distances. This is because any combination of our individual distances will re-use the same prior many times. If stars have large fpus, this product of priors will dominate and introduce a strong bias into the combined distance. This would particularly affect clusters beyond a few kpc.

\subsection{Comparison to other distance estimates}

\begin{figure}
\begin{center}
\includegraphics[width=0.50\textwidth, angle=0]{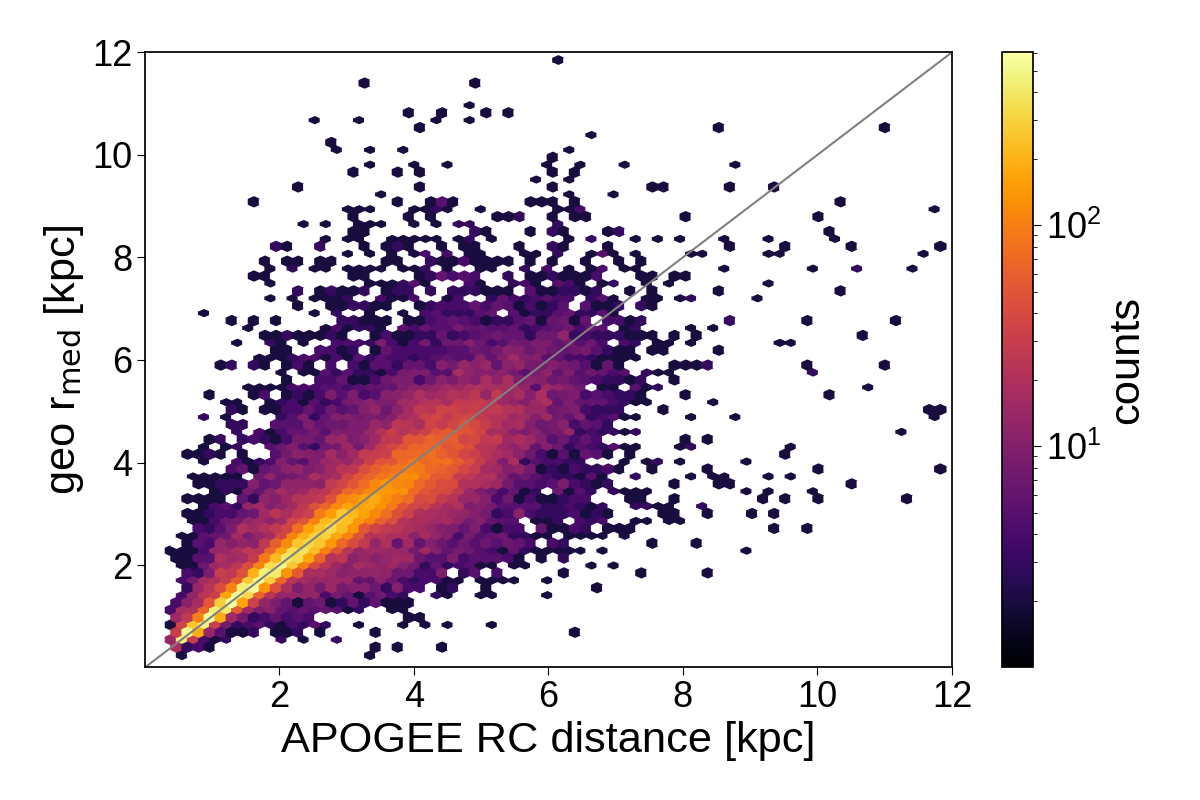}
\includegraphics[width=0.50\textwidth, angle=0]{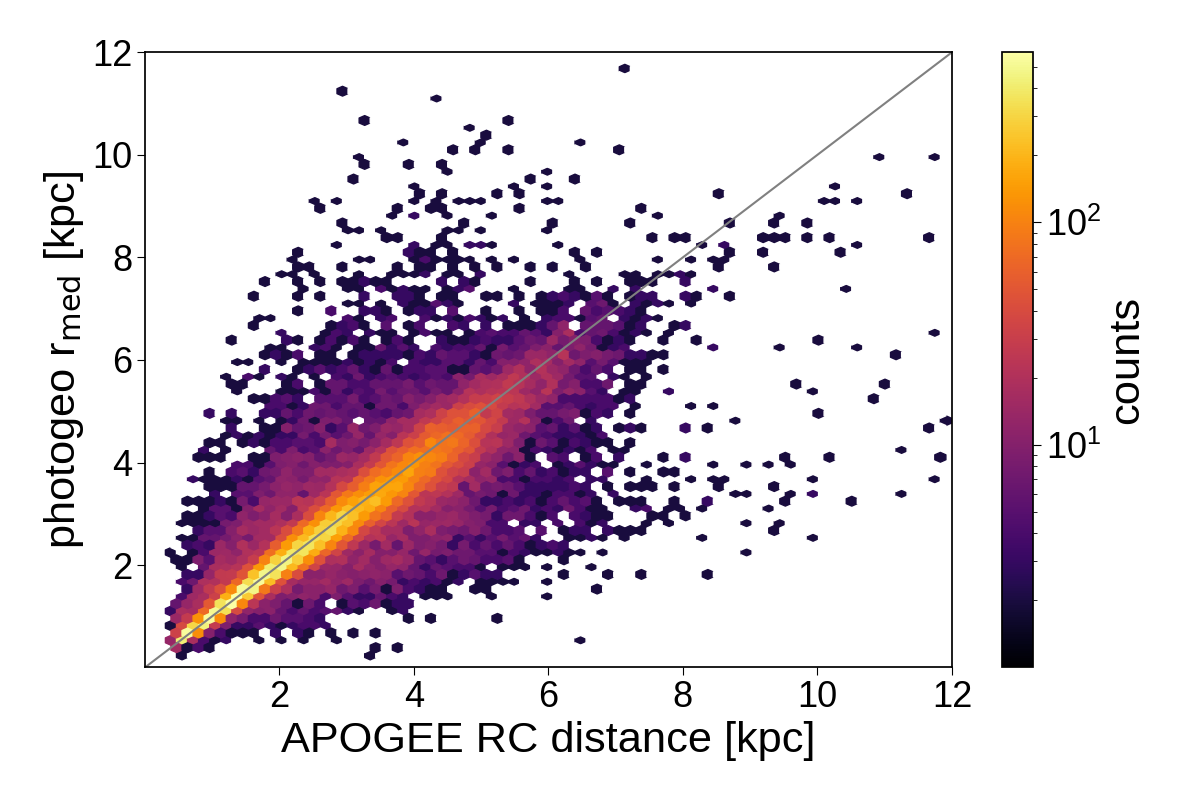}
\caption{Comparison of APOGEE DR16 red clump star distance estimates from \cite{2014ApJ...790..127B}
to our geometric estimates (top panel) and to our photogeometric estimates (bottom panel) for a common sample of 36\,858 sources.
\label{fig:apogee_rc}
}
\end{center}
\end{figure}

Figure~\ref{fig:apogee_rc} compares our distance estimates
for 36\,858 red clump (RC) stars with those estimated 
by \cite{2014ApJ...790..127B} using high-resolution APOGEE \citep{2017AJ....154...94M} DR16 spectra. 
This method selects sources using colour, effective temperature, metallicity, and surface gravity, and is calibrated via stellar evolution models and high-quality asteroseismology data. Given the narrowness of the red clump locus in the parameter space, their distances are expected to be precise to 5\% with a bias of no more than 2\%.

The 5th, 50th, and 95th percentiles of fpu for this sample are 0.01, 0.05, and 0.27 respectively, and of \gmag\ are 10.4, 13.4, and 16.2\,mag respectively. The fractional bias and rms of the deviations of our estimates relative to  those of Bovy et al.\ 
are $+0.05$ and 0.31 respectively for the geometric distances, and $+0.03$ and 0.29 respectively for the photogeometric distances. For reference, the fractional bias and rms of the deviations of the APOGEE red clump estimates relative to the StarHorse \citep{2020A&A...638A..76Q} estimates (see next paragraph) for the same sample are $+0.05$ and 0.21 respectively. The parallaxes for this sample are mostly of such high quality that the prior does not strongly effect our posteriors, although we still see a slight improvement in the photogeometric distances over the geometric ones.
When counting the percentage of sources where the Bovy et al.\ estimate is within our upper and lower bounds (+ 7\% error margin from Bovy et al.) we find that 65\% are compatible with the geometric distances and 69\% with photogeometric (we expect 68\% to be within 1$\sigma$). If we do the same for the StarHorse estimates (which also have upper and lower percentiles) for the red clump sample
we see that 84\% of the StarHorse estimates are within 1$\sigma$ pf the Bovy et al.\ estimates.

\begin{figure}
\begin{center}
\includegraphics[width=0.50\textwidth, angle=0]{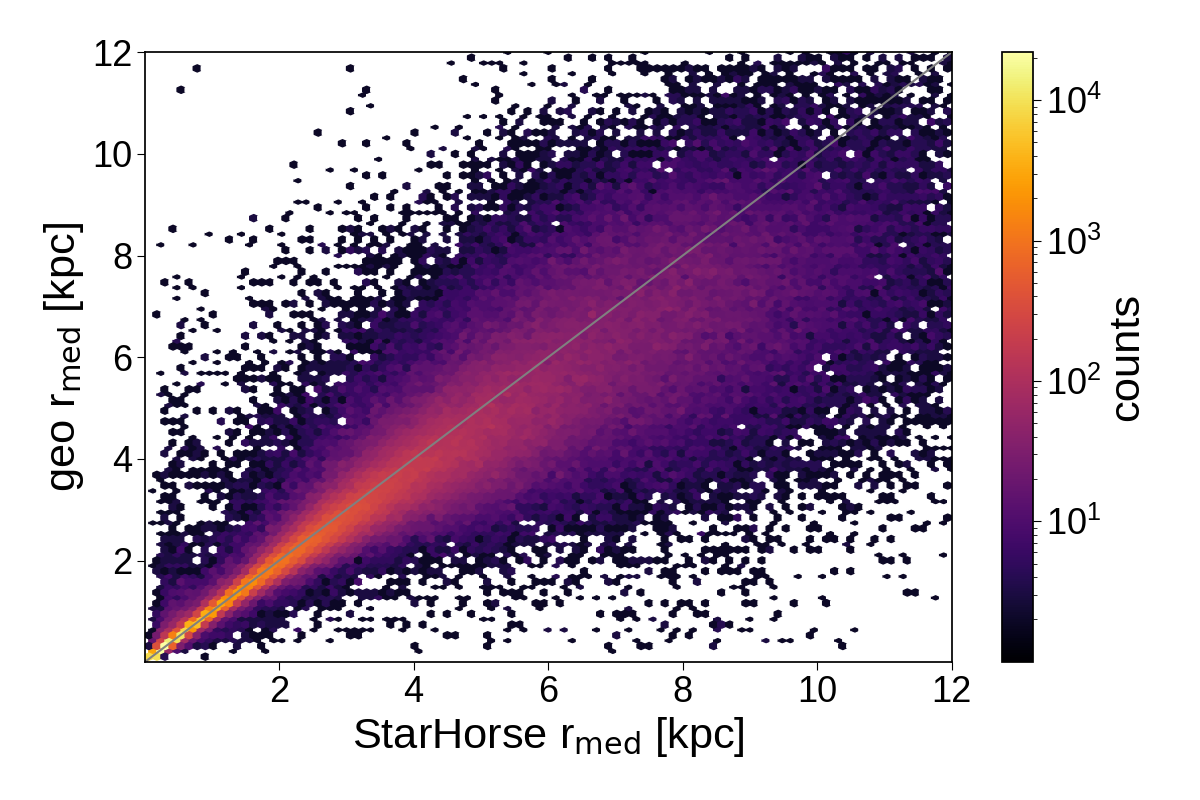}
\includegraphics[width=0.50\textwidth, angle=0]{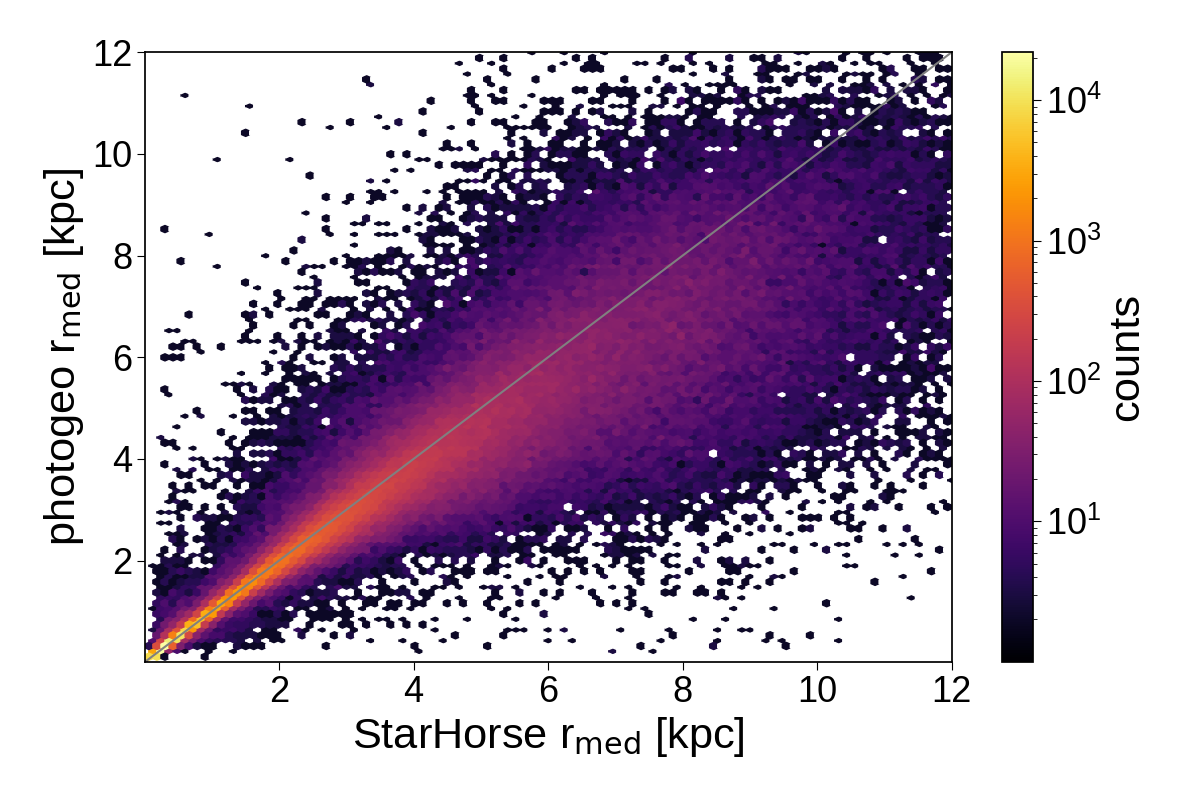}
\caption{Comparison of StarHorse distance estimates from 
\cite{2020A&A...638A..76Q} 
to our geometric estimates (top panel) and to our photogeometric estimates (bottom panel) for a common sample of 307\,105 sources.
\label{fig:starhorse}
}
\end{center}
\end{figure}

Figure~\ref{fig:starhorse} compares our distance estimates for 307\,105 stars with those estimated by \cite{2020A&A...638A..76Q} using their StarHorse method, which uses APOGEE DR16 spectra, multiband photometry, and \gdr{2} parallaxes.
This sample comprises around $1/3$ main sequence stars; the rest are turnoff star and giants, excluding the red clump stars used in the previous comparison. StarHorse estimates a posterior probability distribution which the authors likewise summarize with a median, so our distance estimates are directly comparable. They report achieving typical distance uncertainties of 11\% for giants and ~5\% for dwarfs.

The 5th, 50th, and 95th percentiles of fpu for this sample are 0.002, 0.02, and 0.46 respectively, and of \gmag\ are 10.2, 13.3, and 16.6\,mag respectively. 
The fractional bias and rms  of the deviations of our distance estimates relative to the StarHorse estimates are $0.00$ and 0.30 respectively for the geometric distances, and $-0.01$ and 0.23 respectively for the photogeometric distances. 
As this sample extends to larger distances (and larger fpu) than the sample in Figure~\ref{fig:apogee_rc}, we begin to see that our geometric distances (and to a lesser extent our photogeometric distances) are smaller than the Starhorse distances beyond about 6\,kpc, which is where some of the large fpu sources will have true distances beyond the median of the distance prior.

\vspace*{1em}
\section{Distance catalogue}\label{sec:catalogue}

\subsection{Content}\label{sec:catalogue_content}

\begin{table*}
\begin{center}
\caption{The format of the distance catalogue showing results on five fictitious sources.
The {\tt source\_id} is the same as in \release. 
 {\tt r\_med\_geo}, {\tt r\_lo\_geo}, and {\tt r\_hi\_geo} are the median, 16th percentile, and 84th percentile of the geometric distance posterior in parsec.
{\tt r\_med\_photogeo}, {\tt r\_lo\_photogeo}, and {\tt r\_hi\_photogeo} are the median, 16th percentile, and 84th percentile of the photogeometric distance posterior in parsec. 
Flag is defined in Table~\ref{tab:flag}.
The distances are shown here rounded to three decimal places, but are provided in the catalogue with 32-bit floating point precision, which guarantees a precision of at least 1 part in $2^{24}$ (17 million).
The photogeometric fields can be missing, indicated here with {\tt NA}.
\label{tab:catalogue_extract}
}
\begin{tabular}{rrrrrrrr}
\hline
{\tt source\_id} & {\tt r\_med\_geo} & {\tt r\_lo\_geo} & {\tt r\_hi\_geo} & {\tt r\_med\_photogeo} & {\tt r\_lo\_photogeo} & {\tt r\_hi\_photogeo} & {\tt flag} \\
                 & pc & pc & pc &  pc & pc & pc & \\
\hline
4295806720 & 3547.973 & 2478.490 & 4741.725 & 2705.790 & 2307.170 & 3357.151 & 10033 \\
34361129088 & 291.709 & 275.786 & 306.577 & 290.840 & 277.130 & 304.291 & 10033 \\
38655544960 & 318.105 & 312.888 & 323.334 & 318.807 & 313.264 & 323.045 & 10033 \\
5835726683934945280 & 7547.806 & 4509.953 & 11817.191 & 5299.187 & 4060.932 & 7178.086 & 10033 \\
5835726688222520960 & 6316.000 & 3860.044 & 10591.593 & {\tt NA} & {\tt NA} & {\tt NA} & 10099 \\
\hline
\end{tabular}
\end{center}
\end{table*}

\begin{table}
\begin{center}
\caption{The flag field in the catalogue is a string of five decimal digits ABBCC.
\label{tab:flag}
}
\begin{tabular}{llp{0.3\textwidth}}
\hline
A    &     &  Source magnitude compared to the limit used to make the prior \\
      &  0  &  Source has no G-band magnitude \\
      &  1  &  $\gmag \leq \gmag_{\rm lim}$ \\
      &  2  &  $\gmag > \gmag_{\rm lim}$ \\
\hline
B   &     &  Hartigan dip test for unimodality. 
                  Left digit geometric, right digit photogeometric \\
     &  0  &  unimodal hypothesis okay\\
     &  1  & {\em possible} evidence for multimodality\\
\hline
C   &     &  \QG\ models used in prior. 
                Left digit bluer model, right digit redder model \\
     &  0  &  null (no model)\\
     &  1  &  one-component Gaussian\\
     &  2  &  two-component Gaussian\\
     &  3  &  smoothing spline\\
     & \multicolumn{2}{l}{There is one special setting:} \\
     & 99 & source lacks \gmag\ and/or \bprp \\
\hline
\end{tabular}
\end{center}
\end{table}

The distance catalogue includes an entry for all
1\,467\,744\,818 sources in \release\ that have a parallax.
All of these have geometric distances and 
1\,346\,621\,631 have photogeometric distances.
In comparison there are 1\,347\,293\,721 sources in \release\ that have defined G-band magnitudes\footnote{By this we mean the {\tt phot\_g\_mean\_mag} field is defined. We do not make use of the other estimates of \gmag\ from the \gaia\ catalogue if this field is null.}, \bprp\ colours, and parallaxes, and so could in principle have received a photogeometric distance estimate, but did not due to missing \QG\ prior models. 

The fields in our catalogue are defined in Table~\ref{tab:catalogue_extract}.
3\% of the sources have changed their {\tt source\_id} identifier from  \gdr{2} to \release\ \citep{fabricius_etal_gedr3_validation}, so the {\tt source\_id} cross-match table {\tt dr2\_neighbourhood} provided with \release\ should be used to find the best match before doing source-by-source comparisons between the two releases.
{\tt r\_med\_geo} in Table~\ref{tab:catalogue_extract} is the median (\rmed) of the geometric distance posterior and should be taken as the geometric distance estimate. {\tt r\_lo\_geo} (\rlo) and {\tt r\_hi\_geo} (\rhi) are the 16th and 84th percentiles of the posterior and so together form a 68\% confidence interval around the median. $\rhi - \rmed$ and $\rmed - \rlo$ are therefore both 1$\sigma$-like uncertainties on the distance estimate, and are generally unequal due to asymmetry of the posterior.  
The fields {\tt r\_med\_photogeo}, {\tt r\_lo\_photogeo}, and {\tt r\_hi\_photogeo} are defined in the same way for the photogeometric distance posterior.

We cannot overstate the importance of the uncertainties provided. They reflect the genuine uncertainty in the distance estimate provided by the median. As $\rhi-\rlo$ is a 68\% confidence interval, we expect the true distance to lie outside of this range for a third of the sources. This is the nature of statistical uncertainty and should never be ignored.

The field {\tt flag} is a string of five decimal digits defined in Table~\ref{tab:flag}. 
Flag {\tt A} is set to 2 if the source is fainter than the faintest mock source used to make the prior for that HEALpixel. The estimated distances can still be used. Faint stars tend to have poor parallaxes so the distance uncertainties will generally be larger in these cases.
The two digits of flag {\tt B} refer to the Hartigan dip test, as explained in section~\ref{sec:multimodality}. 
We find that 2\% of geometric posteriors and 3\% of photogeometric posteriors may not be unimodal according to this test, although this test is not particularly accurate, so this is only a rough guide.
Even when the sampled posterior shows a true, significant bimodality (or even multimodality), the 68\% confidence interval sometimes spans all modes. 

The two digits of flag {\tt C} indicate the nature of the two \QG\ models that were used to construct the \QG\ prior. If both numbers are between 1 and 3 then two models bracket the source's colour and were combined by linear interpolation, as explained in section~\ref{sec:QG_prior}. If only one of them is 0 then only a single model was used. 
If both flags are 0 then there is no non-null model within 0.1\,mag  colour of the source, so the photogeometric posterior is not computed. 
There are is one special value of this flag: 
99 means the star lacked the necessary data to compute the photogeometric distance.

We provide additional information on the prior for each HEALpixel in the auxiliary information online, including plots like Figures~\ref{fig:GGDprior_fits}, \ref{fig:mockCQD}, and~\ref{fig:QGmodels1}, and a table with the three parameters of the geometric prior (equation~\ref{eqn:GGDprior}).

\subsection{Filtering}

We have not filtered out any results from our catalogue. Parallaxes with spurious parallaxes remain, as do sources with negative parallaxes (the latter is no barrier to inferring a sensible distance; \citealt{2015PASP..127..994B}).
Any filtering should be done with care, as it often introduces sample biases.
The flag field we provide is for information purposes; we do not recommend to use it for filtering. Lower quality distances will arise from lower quality input data.
These can be identified using the various quality fields in the main Gaia catalogue of \release, which is easily cross-matched to our catalogue using the {\tt source\_id} field, as shown in the example in section~\ref{sec:access}. Useful quality metrics may be {\tt ruwe}, {\tt parallax\_over\_error}, and {\tt astrometric\_excess\_noise}, as defined in the \release\ documentation, where users will also find advice on their use.
See in particular section 3.2 of
\cite{fabricius_etal_gedr3_validation} for suggestions for filtering spurious parallaxes.

Parallaxes from the 6p astrometric solutions (identified by {\tt astrometric\_params\_solved = 95}) are not as accurate as those from the 5p solutions \citep{lindegren_etal_gedr3_astrometry}, because they were normally used in more problematic situations, such as crowded fields, and are also fainter on average than the 5p solutions. Sources with 6p solutions should not be automatically removed, however. Their larger parallax uncertainties reflect their lower quality.
In some applications users may want to filter out sources with large absolute or relative distance uncertainties.  One must exercise caution here, however, because uncertainty generally correlates with distance and/or magnitude (among other things), so filtering on these quantities will introduce sample biases. 

\subsection{Use cases}

For stars with positive parallaxes and $\fpu < 0.1$, the inverse parallax is often a reasonably good distance estimate for many purposes (when using a suitable parallax zeropoint). This applies to 98 million sources in \release. For sources with negative parallaxes or $\fpu > 1$ (704 million sources), our distances will generally be prior dominated, and while the photogeometric distances could still be useful, the geometric ones are probably less so. The sweet spot where our catalogue adds most value is for the remaining 665 million sources with $0.1 < \fpu < 1$.

The choice of whether to use our geometric or photogeometric distance depends on the specific situation and what assumptions you are willing to accept.
In the limit of negligible parallax uncertainties they will agree. 
At large fractional parallax uncertainties our photogeometric distances will generally be more precise than geometric ones, because they use more information and have a stronger prior (see Figures~\ref{fig:real_inference_1} and~\ref{fig:real_inference_2}). Whether they are also more accurate depends on how well the \QG\ prior matches to the true (but unknown) \QG\ distribution. The \QG\ model reflects the stellar population and interstellar extinction in a small patch of sky (HEALpixel of area 3.36\,sq.~deg). The \mock\ catalogue and our prior should model these reasonably well at  higher Galactic latitudes, but may be less accurate at lower latitudes where extinction is higher and the stellar populations along the line-of-sight are more complicated. If you do not want to rely on colour and magnitude information in the distance inference, use the geometric distance, as the distance prior is less sensitive to the exact stellar population in \mock. 

Some example use cases are as follows.
\begin{enumerate}

\item Look-up of distance (or distance modulus) for particular sources of interest using their {\tt source\_id} or other identifier matched to this. \release\ includes a crossmatch to many existing catalogues. Positional crossmatches can also be done on the \release\ data site or using TAP uploads, and at other sites that host our catalogue.

\item Identification of sources within a given distance (or distance modulus) range. The confidence intervals should be used to find all sources with a distance \dist\ satisfying $k(\rmed - \rlo) < \dist < k(\rhi - \rmed)$, where the size of $k$ will depend on the desired balance between completeness and purity of the resulting sample. A better approach would be to use the actual posterior to get a probability-weighted sample. For the geometric distances our posterior can be reconstructed using the geometric distance prior provided for each HEALpixel in the auxiliary information online. Readers interested in using our photogeometric priors should contact the authors.

\item Construction of absolute-colour-magnitude diagrams. One of the reasons that we provide quantiles for our distance estimates is that $5\logten(\rmed)-5$ is the median of the distance modulus posterior. (This would not be the case if we provided the mean or mode, for example.) Using \gmag\ from \release\ one can then compute \QG, and from this the absolute magnitude \MG, if the extinction is zero or otherwise known. The same can be done for any photometric band from any other catalogue. When computing \QG\ in this way with equation~\ref{eqn:continuity}, the user should remember to apply the correction to the \release\ G-band magnitude as described in section 8.3 of \cite{riello_etal_gedr3_photometry}.

\item For constructing the three-dimensional spatial distribution of stars in some region of space. This may also assist selection of candidates in targeted follow-up surveys.

\item As a baseline for comparison of distance or absolute magnitude estimates obtained by other means.

\item Our distances could be used for another layer of inference, such as computing transverse velocities using also the \release\ proper motions, although users will need to consider the appropriate error propagation. In particular, if the error budget is not dominated by a single source (e.g.\ not just the distance),  users are advised to infer their desired quantities directly from the original parallaxes, perhaps using the priors provided here.
\end{enumerate}

Users should realise that uncertainties in the parallaxes in \release\ are correlated between different sources to a greater or less degree depending on their angular separations \citep{lindegren_etal_gedr3_astrometry, fabricius_etal_gedr3_validation}. Caution must therefore be exercised when combining either the parallaxes or our distances, e.g.\ averaging them to determine the distance to a star cluster. In such a case the simple ``standard error in the mean" may underestimate the true uncertainty, and the same prior would be used multiple times. One should instead set up a joint likelihood for the sources that accommodates the between-source correlations and solve for the cluster distance directly.

\subsection{Access}\label{sec:access}

Our distance catalogue is available from the German Astrophysical Virtual Observatory at
\url{http://dc.g-vo.org/tableinfo/gedr3dist.main} where it can be queried via TAP and ADQL. 
This server also hosts a reduced version of the main Gaia \release\ catalogue (and \mock). Typical queries are likely to involve a join of the two catalogues. By way of example, the following query returns coordinates, our distances, \bprp, and the two \QG\ values using the median distances, for all stars with a low {\tt ruwe} in a one-degree cone in the center of the Pleiades.
This should run in about one second and return 22\,959 sources.

\small{
\begin{verbatim}
SELECT 
  source_id, ra, dec,
  r_med_geo, r_lo_geo, r_hi_geo,
  r_med_photogeo, r_lo_photogeo, r_hi_photogeo,
  phot_bp_mean_mag-phot_rp_mean_mag AS bp_rp,
  phot_g_mean_mag-5*LOG10(r_med_geo)+5 AS qg_geo,
  phot_g_mean_mag-5*LOG10(r_med_photogeo)+5 
    AS gq_photogeo
FROM gedr3dist.main
  JOIN gaia.edr3lite USING (source_id)
WHERE ruwe<1.4 
  AND DISTANCE(ra, dec, 56.75, 24.12)<1
\end{verbatim}
}

A bulk download for the catalogue is also available at the URL given above. Our catalogue will also become available soon
together with the full \release\ catalogue hosted at
\url{https://gea.esac.esa.int/archive/} and its partner data centers.
At these sites the table names \verb|gedr3dist.main| and \verb|gaia.edr3lite| may well be different.

\subsection{Limitations}

When using our catalogue users should be aware of its assumptions and limitations.
\begin{enumerate}

\item We summarize the posteriors using only three numbers (quantiles), which cannot capture the full complexity of these distributions. This is more of a limitation for the photogeometric posteriors. The confidence intervals should not be ignored.

\item Most sources in \release\ have large fractional parallax uncertainties and our distances correspondingly have large fractional uncertainties, especially for the geometric distances.

\item The poorer the data, the more our prior dominates the distance estimates. Our prior is built using a sophisticated model of the Galaxy that includes 3D extinction, but it will not be perfect. If the true stellar population, extinction, or reddening law are very different in reality, our distances will be affected. In section~\ref{sec:mock_qualitative_analysis} we explained, using results on simulated data, what biases can occur and why. 

\item Sources with very large parallax uncertainties will have a posterior dominated by the prior.
The median of this varies between 745 and 7185\,pc depending on HEALpixel (Figure~\ref{fig:GGDprior_median_sky}). Stars with large fpus that truly lie well beyond the prior's median will have their geometric distances underestimated; stars with large fpus that lie closer than the prior's median will have their geometric distances overestimated. As distant stars generally have larger fpu than nearby stars, and distant stars are more numerous, the former characteristic will dominate among poor quality data. This leads to a bias in distance estimates, one that is probably unavoidable (see appendix~\ref{sec:thoughts_prior}). Poor data remain poor data.

\item Our prior is spatially discretized at HEALpixel level 5, i.e.\ in patches of 3.36\,sq.~deg.\ on the sky. The distance prior and CQD change discontinuously between HEALpixels, and this may be visible in sky maps of posterior distances.  The \QG\ priors (constructed from the CQD) are formed by a linear interpolation over colour whenever possible, so in these cases there should be no discontinuity of distance with colour within a HEALpixel.

\item Our inferred distances retain all of the issues affecting the parallaxes, some of which have been explored in the \release\ release papers \citep{lindegren_etal_gedr3_astrometry, fabricius_etal_gedr3_validation}.
We applied the parallax zeropoint correction derived by \cite{lindegren_etal_gedr3_parallaxzp}, which is better than no correction or a single global correction, but is not perfect. Any error in this will propagate into our distance estimates.
The published parallax uncertainties are also probably also underestimated to some degree \citep{fabricius_etal_gedr3_validation}.
\cite{gedr3_release} and \cite{riello_etal_gedr3_photometry}
report some issues with the \release\ photometry, 
such as biased BP photometry and therefore \bprp\ colours for very faint sources,
which could affect our photogeometric distances.
These distance estimates additionally suffer
from any mismatch between the published \release\ photometry and the modelling of this 
-- in particular the passbands -- used in the \mock\ catalogue, which forms the basis for our \QG\ priors.\footnote{We compared 
simulations of the G-band magnitude and the BP-RP colour
between the \mock\ passbands and those published for \release,
using isochrones at 4\,Myr and 1\,Gyr.
The differences in the G magnitudes are below 6\,mmag, except for sources bluer than $-0.15$, where it can be as high as 700\,mmag. For \bprp\ using the BP bright band (in \mock), the difference is around 10\,mmag, but up to 25\,mmag for sources with $\bprp > 1.2$\,mag and up to 100\,mmag for sources with $\bprp < -0.2$\,mag. For BP faint, the \bprp\ difference is around 20\,mmag, but up to 60\,mmag
for sources with $\bprp > 0.5$\,mag and up to 100\,mmag for sources with $\bprp < -0.15$\,mag.}
Note that we applied the G-band magnitude correction to the \release\ photometry as described by \cite{riello_etal_gedr3_photometry}.

\item We implicitly assume that all sources are single stars in the Galaxy. 
Our distances will be incorrect for extragalactic sources. The geometric distances will be wrong for unresolved binaries if the parallax for the composite source is affected by the orbital motion. Even when this is not the case the photogeometric distance may still be wrong, because the G-band magnitude will be brighter than the \QG\ prior expects (binaries were not included in the prior).

\item By design we infer distances for each source independently. If a set of stars is known to be in cluster, and thus have a similar distance, this could be exploited to infer the distances to the individual stars
  more accurately than we have done here.
  In its most general form this involves a joint inference over multiple sources.
Various methods exist in the literature for doing this, such as \cite{2014A&A...564A..49P}, \cite{2018A&A...618A..93C}, and \cite{2020arXiv201000272O}.
Likewise, in order to estimate the distance to the cluster as a whole, one should be aware that averaging our individual distances
will compound the prior. If the fpu of the individual sources is large, this product of priors would dominate the distance estimate more than desired. A joint inference can easily be set up overcome this.
  
\end{enumerate}

\vspace*{1em}
\section{Summary}\label{sec:summary}

We have produced a catalogue of geometric distances for 1.47 billion stars and photogeometric distances for 92\% of these. These estimates, and their uncertainties, can also be used as estimates of the distance modulus. Geometric distances use only the \release\ parallaxes. Photogeometric distances additionally use the \gmag\ magnitude and \bprp\ colour from \release.
Both types of estimate involve direction-dependent priors constructed from a sophisticated model of the 3D distribution, colours, and magnitudes of stars in the Galaxy as seen by \gaia, i.e.\ accommodating both interstellar extinction and a \gaia\ selection function.
Tests on mock data, but moreover validation against independent estimates and open clusters, suggest our estimates are reliable out to several kpc. 
For faint or more distant stars the prior will often dominate the estimates. We have identified various use cases and limitations of our catalogue.

Our goal has been one of inclusion: to provide distances to as many stars in the \release\ catalogue as possible. This has required us to make broad, general assumptions. If one focuses on a restricted set of stars with some approximately known properties, it will be possible to construct more specific priors, and to use these to infer more precise and more accurate distances. Better distances may also be achievable by using additional data, such as spectroscopy or additional photometry.

\acknowledgments
We thank the IT departments at MPIA and ARI for computing support.
This work was funded in part by the DLR (German space agency) via grant 50 QG 1403.
It has made use of data from the European Space Agency (ESA) mission \gaia\ (\url{http://www.cosmos.esa.int/gaia}), processed by the \gaia\ Data Processing and Analysis Consortium (DPAC, \url{http://www.cosmos.esa.int/web/gaia/dpac/consortium}). Funding for the DPAC has been provided by national institutions, in particular the institutions participating in the \gaia\ Multilateral Agreement. 
This research made use of: TOPCAT, an interactive graphical viewer and editor for tabular data \citep{2005ASPC..347...29T}; Vaex, a tool to visualize and explore big tabular data \citep{2018A&A...618A..13B}; matplotlib, a Python graphics library \citep{Hunter:2007}; \texttt{HEALpix} \citep{2005ApJ...622..759G} and \texttt{healpy} \citep{2019JOSS....4.1298Z}; the NASA Astrophysics Data System; the VizieR catalogue access tool, CDS, Strasbourg.

\facility{Gaia}


\appendix

\section{Thoughts on a better distance prior}\label{sec:thoughts_prior}

The strong dependence of the geometric posterior on the distance prior in the limit of large parallax uncertainties is an unavoidable consequence of inference with noisy data. We saw something similar in paper IV. 
This leads to a distance bias mostly for distant stars with large fpu. Could this be avoided? Conceptually one would like a distance prior that depends on the true fpu, but this is impossible because the true parallax is not known. One may be tempted to use the measured fpu instead, but this is not what we want: a star with a large true fpu could have a small measured fpu due to noise, and thereby be treated incorrectly. Its use is also be theoretically dubious because it places the parallax -- a measurable -- in the prior, as well as in the likelihood. 
We experimented with using a prior conditioned on \sigparallax, but found that this did not help (see the technical note GAIA-C8-TN-MPIA-CBJ-089 with the auxiliary information online).
One may achieve something close to what is desired by simply shifting the distance prior to greater distances, so that it better represents stars with a larger true fpu, which is where the prior is needed more. 
Yet this would detrimentally affect the distance estimates for nearby stars. It seems a poor trade-off to sacrifice accuracy on high-quality data for a better prior on low-quality data. 
Conditioning the prior on the star's magnitude may help, and this is what our photogeometric distances do (section~\ref{sec:photogeometric_distance}).

\section{The limit of poor parallaxes}\label{sec:poor_parallax_limit}

We tend to think that a large fpu means that the likelihood is uninformative and that the posterior converges towards the prior. Consider a red clump star in the LMC with a true parallax of 0.02\,mas and a typical parallax uncertainty of 0.2\,mas for a star with $\gmag = 19$\,mag. The true fpu is $0.2/0.02=10$. Let's assume initially that we actually measure a parallax of 0.02\,mas, i.e.\ we have an measured fpu of 10.
(Of course in this lucky case the inverse parallax would be the correct distance, but it's very rare in practice.)
In the LMC HEALpixel 8275 our distance prior has a median of 1.2\,kpc because we exclude the LMC from our prior, so we might expect to see many sources with this inferred distance. In fact we see many sources with larger inferred distances (see the plot with the auxiliary online information). The reason is that the likelihood of a measurement of 1\,mas (corresponding to a distance of 1\,kpc) is still at 4.9\,$\sigma$ and therefore quite unlikely. This shows that even when the fpu is large the parallax can be quite informative. 

One should remember, however, that our inference never sees the true parallax but only the measured parallax, which is normally distributed around the unknown true value (with a standard deviation which is also only estimated). So it is quite likely that our measurement of the above red clump star gives us a parallax measurement of, say, 0.4\,mas. In that case the measured fpu is 0.5 and the likelihood of 1\,mas, i.e.\ a 1\,kpc distance, is only 3$\,\sigma$ away from this measurement. Taking the parallax \emph{measurement} into account essentially redistributes probability mass into the wings of the likelihood and therefore to higher and lower (also negative) parallax values. Given the truncation of negative parallaxes when calculating the posterior, this implies that the median distance estimate is lower for the true measurements, compared to the idealised inference using the true parallax.
Similarly, one should be careful not to interpret plots involving the measured fpu as though it were the true fpu.

\bibliographystyle{aasjournal}
\bibliography{distances,gaia}

\end{document}